\documentclass[aps, prl reprint, twocolumn, noeprint, longbibliography]{revtex4-2}
\usepackage{amsmath, bm, amssymb, dsfont, wasysym}  % Math
\usepackage{indentfirst}  % Indent 1st paragraph after Section
\usepackage{graphicx}  % Figures
\usepackage{color}  % Colors, mainly for markup
\usepackage{hyperref}  % Links
\usepackage{ulem}  % Underlining
\usepackage{changes}
\def\etal{\textit{et al.}}

\def\Rb{$^{87}\mathrm{Rb}$}  % Rubidium-87
\def\mf{m_{F}}  % Magnetic sublevels
\def\ket#1{\left|#1\right\rangle}  % Dirac notation ket
\def\braket#1{\left\langle #1 \right\rangle} % Diract notation braket
\def\wilson{\mathcal{W}}  % Wilson loop
\def\identity{\hat{\mathds{1}}}  % Identity operator
\def\holonomy#1{\hat{\Gamma}_{A}(#1)}  % Holonomy
\hypersetup{
    colorlinks=true,
    linkcolor=blue,
    citecolor=blue,
    filecolor=blue,
    urlcolor=blue,
}

\begin{document}

% -----------------------------------
% ----- Preamble -----
% -----------------------------------
\title{Investigation of Floquet engineered non-Abelian geometric phase for holonomic quantum computing}

% Authors/Affiliations
\author{Logan W. Cooke$^1$}
\author{Arina Tashchilina$^1$}
\author{Mason Protter$^1$}
\author{Joseph Lindon$^1$}
\author{Tian Ooi$^1$}
\author{Frank Marsiglio$^{1,2}$}
\author{Joseph Maciejko$^{1,2}$}
\author{Lindsay J. LeBlanc$^1$}
\affiliation{$^1$Department of Physics, University of Alberta, Edmonton, Alberta, Canada}
\affiliation{$^2$Theoretical Physics Institute, University of Alberta, Edmonton, Alberta, Canada}

% Date
\date{January 16, 2024}

% --- Abstract ---
\begin{abstract}
    Holonomic quantum computing (HQC) functions by transporting an adiabatically degenerate manifold of computational states around a closed loop in a control-parameter space; this cyclic evolution results in a non-Abelian geometric phase which may couple states within the manifold. Realizing the required degeneracy is challenging, and typically requires auxiliary levels or intermediate-level couplings. One potential way to circumvent this is through Floquet engineering, where the periodic driving of a nondegenerate Hamiltonian leads to degenerate Floquet bands, and subsequently non-Abelian gauge structures may emerge. Here we present an experiment in ultracold $^{87}$Rb atoms where atomic spin states are dressed by modulated RF fields to induce periodic driving of a family of Hamiltonians linked through a fully tuneable parameter space. The adiabatic motion through this parameter space leads to the holonomic evolution of the degenerate spin states in $SU(2)$, characterized by a non-Abelian connection. We study the holonomic transformations of spin eigenstates in the presence of a background magnetic field, characterizing the fidelity of these single-qubit gate operations. Results indicate that while the Floquet engineering technique removes the need for explicit degeneracies, it inherits many of the same limitations present in degenerate systems.
\end{abstract}

% Compile
\maketitle

% -----------------------------------
% ----- Add/organize sections -----
% -----------------------------------
\section{Introduction}
\label{s:introduction}

Quantum computing promises to solve some classically hard problems more efficiently than conventional (classical) methods, but both coherent and incoherent noise pose real barriers to practical deployment and use~\cite{DiVincenzo.2000}. Designing better qubits or error-correcting codes is a significant area of research~\cite{Devitt.2013}, as is the search for new fault-tolerant quantum control techniques.

Holonomic quantum computing (HQC)~\cite{Pachos.2001, Zanardi.1999qb, Pachos.2000} is a promising approach that uses geometric phase in contrast to the more conventional gates that rely on dynamical phase. Geometric phases are independent of any details in the control Hamiltonian, instead depending only on the curvature in a state's Hilbert space as it varies with a set of control parameters. Geometric gates have long been thought to host intrinsic fault-tolerance when compared to dynamical gates, though recent work suggests that fault-tolerance does not depend on the type of phase, but rather on the details of the control Hamiltonian itself~\cite{Colmenar.2022}.

In HQC, computational states are encoded into a degenerate subspace of the control Hamiltonian. The Hamiltonian is varied adiabatically in a cyclic manner; states evolve and mix according to a non-Abelian connection in the parameter space. The evolution operator is referred to as a holonomy\footnote{Also referred to as an \emph{anholonomy} in earlier work on geometric phase.}, due to its geometric interpretation~\cite{Pachos.2001, Zanardi.1999qb, Pachos.2000, berry1984quantal}. This protocol necessitates robust degeneracies, which are typically acquired through coupling to auxiliary levels~\cite{Zhang.20212cna}. HQC protocols are also generally seen as slow due to the adiabatic criterion~\cite{Zhang.20212cna}. As such, there are non-adiabatic generalizations~\cite{Sjöqvist.2012} in which the role of degeneracy is relaxed, but the dynamical contributions to the phase that arise from this and the breaking of the adiabaticity impose strict conditions on the details of each gate. There have been several experimental demonstrations of both conventional and non-adiabatic HQC in trapped ions~\cite{Duan.2001, Ai.2022}, neutral atoms~\cite{Shui.2021, Leroux.2018}, liquid nuclear-magnetic resonance systems~\cite{Feng.2013, Jones.2000}, Rydberg atoms~\cite{Xu.2022q69}, solid-state systems~\cite{Yan.2019, Li.2022, Yang.2023, Jr.2013}, photonics devices~\cite{Pinske.2020, Kremer.2019}, and nitrogen-vacancy centers~\cite{Sekiguchi.2017, Nagata.2018, Zu.2014, Arroyo-Camejo.2014}.

Topological quantum computing (TQC) also relies on non-Abelian holonomies, but differs from HQC. In TQC, quantum information is encoded in the phase of non-Abelian anyons [26, 27]; these are particles that obtain a non-Abelian geometric phase under exchange, in contrast to fermions and bosons which only obtain \emph{Abelian} phases of $\pi$ and $0$ under exchange, respectively. Qubits may be encoded onto a degenerate manifold of multi-anyon states, and gates are performed by moving anyons around each other, called braiding. In this case, the unitary transformation within this manifold depends only on the topological character of the path, that is, whether the anyon's path encapsulates another anyon or not. In this way the phase is still described by a non-Abelian holonomy, but is less sensitive to the detailed geometry of the adiabatic path than in HQC.

Recently, Floquet engineering was proposed as a path to produce robust degeneracies in systems which are otherwise non-degenerate~\cite{Novičenko.2019, Novičenko.2017}. Through periodic modulation of a control Hamiltonian, degeneracies occur regardless of the underlying energetic structure. As such, this method may be readily applied as an HQC scheme without the need for any auxiliary levels or excited state couplings, with proposed implementations in ultracold neutral atoms~\cite{Chen.2020} and Rydberg atoms~\cite{Wang.2021}. Additionally, for quantum simulation, this technique provides a path forwards to realizing interesting non-Abelian artificial gauge fields~\cite{Galitski.2019, Sugawa.2021}.

Here, we present a proof-of-concept experimental investigation of Floquet engineered single-qubit holonomic gates in an optically trapped ensemble of $~^{87}$Rb. The holonomies investigated here may be readily implemented in other platforms. We perform several primitive gates and, through the tomographic reconstruction of the holonomies, we report their fidelities. Measurements are made in the presence of drifting background magnetic fields, the impact of which would normally be negligible over the time scales of each gate; however, due to the dynamics introduced by the Floquet driving this uncontrolled background has a substantial impact on the holonomy and its geometric nature. We quantify this impact and discuss the implications it has on the practicality of this protocol and its fault-tolerance, both in the cold-atom context, and more generally for generic Floquet-driven platforms.

\section{Theory of Floquet-Engineered Degeneracy}
\label{s:theory}

In this work, we consider a system of $N = 2F+1$ spin levels, coupled by an external field~\cite{Chen.2020}, which applies to any platform with quantized spin or pseudospin levels. Coupling is only between levels adjacent in energy, and thus the Hamiltonian is expressed in terms of the vector of $N \times N$ spin matrices, $\bm{\hat{F}}$. The driving field's amplitude, frequency, and phase are periodically modulated, such that we obtain the Hamiltonian (in units where $\hbar=1$),
\begin{gather}
    \hat{H}(t) = \hat{V}(t) \cos\omega t \label{eq:mod-ham}, \\
    \hat{V}(t) = \Omega_0 \, \bm{q}(t) \cdot \bm{\hat{F}}. \label{eq:mag-ham}
\end{gather}
Equation~\ref{eq:mag-ham} is the Hamiltonian for a spin in a magnetic field with magnitude $\Omega_0$, where the direction of the field is defined by the unit vector $\bm{q}=\left( \sin\Theta\cos\Phi, \sin\Theta\sin\Phi, \cos\Theta \right)^{\intercal}$. This fictitious field is controlled through the phase of the driving field $\Phi(t)$ and control parameter $\Theta(t)$, which relates to the amplitude/frequency modulation envelope. Hence, the control parameter space is the unit sphere spanned by the angles $\left\{ \Theta(t), \Phi(t) \right\}$. For a full derivation of this Hamiltonian, see Appendix~\ref{s:supp-RF}. The entire Hamiltonian $\hat{V}(t)$ is periodically driven with Floquet frequency $\omega$.

\begin{figure}
    \includegraphics{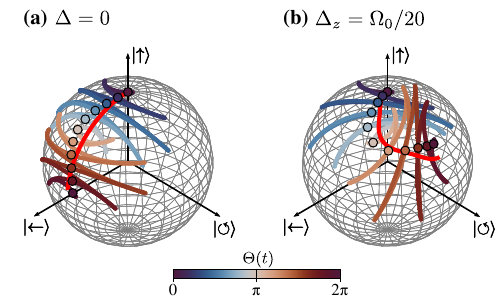}
    \caption{Spin-1/2 simulations of state trajectories on the Bloch sphere (see Appendix~\ref{s:supp-numerics} for details), with axes indicted by the $-1$, $-1$, and $+1$ eigenstates of $\hat{\sigma}_x$, $\hat{\sigma}_y$, and $\hat{\sigma}_z$ respectively. In both (a) and (b), $\Omega_0 / \omega = 1$, and $\Omega = \Omega_0 / 10$. Trajectories are shown for loop $\ell_1$ (see Table~\ref{tab:loops}), with detuning (Eq.~\ref{eq:delta-ham}) $\Delta_z = 0$ (a), and $\Delta_z = \Omega_0 / 20$ (b). Colored lines show trajectories calculated in the rotating frame (Eq.~\ref{eq:mag-ham}) with colors denoting progress through the loop, and points (circles) sampled stroboscopically, at each half-integer period of the Floquet driving frequency $\omega$. The solid (red) line is simulated in the Floquet basis. The disagreement between the stroboscopic points and Floquet-basis simulations are due to non-adiabatic corrections, which for the parameters chosen here amounts to about a 3\% error, in terms of the fidelity (Eq.~\ref{eq:fidelity}); as $\Omega / \omega \rightarrow 0$ the evolution becomes more adiabatic.}
    \label{fig:level-scheme}
\end{figure}

Assuming that the timescales over which $\bm{q}(t)$ changes are very long compared to $2\pi/\omega$, we treat $\hat{H}(t)$ as nearly periodic and transform into the Floquet basis via the unitary operator $\hat{U} = \exp [i \hat{V}(t) \sin (\omega t) / \omega  ]$, usually called the micromotion operator. Following this we restrict our attention to the zeroth Floquet band~\cite{Chen.2020, Novičenko.2017, Novičenko.2019}, where the Hamiltonian becomes (see Appendix~\ref{s:supp-floquet}),
\begin{gather}
    \hat{H}_{\mathrm{Floq.}} = {\partial \bm{q} \over \partial t} \cdot \bm{\hat{A}} \label{eq:floq-ham}, \\
    \bm{\hat{A}} = g \bm{\hat{F}} \times \bm{q} \label{eq:conn},
\end{gather}
where $g = 1 - J_0 \left( \Omega_0 / \omega \right)$, and $J_0$ is the zeroth-order Bessel function  of the first kind. This Hamiltonian only depends on changes to $\bm{q}$; for any static choice of $\bm{q}$ (point in parameter space) the Hamiltonian is zero, hence the Floquet states are trivially degenerate. The Schr\"{o}dinger equation for Eq. (\ref{eq:floq-ham}) can be written as
\begin{align}
\bm{\nabla}_{\bm{q}} \ket{\psi} = -i \bm{\hat{A}}(\bm{q}) \ket{\psi},
\label{eq:geo-schrod}
\end{align}
where $\bm{\nabla}_{\bm{q}}$ is the gradient in the parameter space of $\bm{q}$. Equation~\ref{eq:geo-schrod} is a purely geometric equation describing the parallel transport of a state $\ket{\psi}$ in the parameter space of $\bm{q}$ with non-Abelian connection $\bm{\hat{A}}(\bm{q})$~\cite{Wilczek.1984}. Changing $\bm{q}$ along some path in parameter space results in the accumulation of a fully geometric phase determined by the connection  $\bm{\hat{A}}$. As $\bm{q}$ changes in time, the Hamiltonian takes on instantaneous eigenvalues proportional to, but smaller than, the rate of change. As such, the system remains adiabatically degenerate, and its evolution will depend only on the geometry of the path traced out, and not on any dynamical details.

As is typical in HQC, we restrict our attention to the evolution of the system over loops in parameter space; we thus assume that $\bm{q}(t)$ varies over a cyclic path, $\ell$ in a period $T = 2\pi / \Omega$. According to Eq.~\ref{eq:geo-schrod} the evolution operator over a path $\ell$ is the holonomy~\cite{Zanardi.1999qb, Pachos.2001},
\begin{equation}
    \hat{\Gamma}_{A}(\ell) = \mathcal{P} \exp \left[ -i \oint_{\ell} d\bm{q} \cdot \bm{\hat{A}}(\bm{q}) \right],
    \label{eq:holonomy}
\end{equation}
where $\mathcal{P}$ is the path-ordering operator. Several loops are explored here, summarized in Table~\ref{tab:loops}. Regardless of the spin manifold $F$, these loops generate transformations in $SU(2)$; as shown in Table~\ref{tab:loops}, the loops $\ell_{1-3}$ produce spin rotations by angle $2\pi g$ about the $x$, $y$, or $z$ axes. As such, the results shown here easily generalize to universal single qubit gate operations. The loops $\ell_{4-6}$ demonstrate how other phases may be generated.

\begin{table}
\caption{The various loops considered here parameterized by $\Theta$ and $\Phi$, and the corresponding holonomies. Evolution occurs over a single period of $\Omega = \Omega_0/10$. The appearance of all three spin matrices in the phase demonstrates how these loops generate transformations in $SU(2)$. The measured average fidelities (Eq.~\ref{eq:fidelity}) without ($\bar{\mathcal{F}}$) and with ($\bar{\mathcal{F}}^{\Delta}$) detuning (Eq.~\ref{eq:delta-ham}) included in the fitting model are given below. For the ratio $\Omega_0/\omega=1$ used in the experiment, $g=1-J_0(\Omega_0/\omega)\approx0.23$. Note that a closed-form solution for the holonomies $\hat{\Gamma}_{A} (\ell_5)$ and $\hat{\Gamma}_{A} (\ell_6)$ are unknown because $\bm{\hat{A}}$ does not commute at different points on these paths.}
\begin{ruledtabular}
\begin{tabular}{ c c c c c c }
    Loop & $\Theta(t)$ & $\Phi(t)$ & $\hat{\Gamma}_{A}(\ell)$ & $ \bar{\mathcal{F}}$ & $ \bar{\mathcal{F}}^{\Delta}$ \\ \hline
    $\ell_1$ & $\Omega t$ & $0$ & $\exp\left( -i 2\pi g \hat{F}_y \right)$ & 0.43 & 0.75 \\
    $\ell_2$ & $\Omega t$ & $\pi / 2$ & $\exp\left( i 2\pi g \hat{F}_x \right)$ & 0.51 & 0.82 \\
    $\ell_3$ & $\pi / 2$ & $\Omega t$ & $\exp\left( -i 2\pi g \hat{F}_z \right)$ & 0.34 & 0.82 \\
    $\ell_4$ & $\Omega t$ & $\pi / 4$ & $\exp\left[ i \sqrt{2}\pi g \left(\hat{F}_x - \hat{F}_y\right) \right]$ & 0.62 & 0.90 \\
    $\ell_5$ & $\pi / 4$ & $\Omega t$ & - & 0.47 & 0.87 \\
    $\ell_6$ & $\Omega t$ & $\Omega t$ & - & 0.53 & 0.79 \\
\end{tabular}
\end{ruledtabular}
\label{tab:loops}
\end{table}

The rate at which loops are traversed, $\Omega$, is constrained by the adiabatic condition~\cite{Novičenko.2019}, $\Omega \ll \omega$. If we choose $\Omega$ to be a sub-harmonic of $\omega$, then, at the end of a loop (and every half-integer period of $\omega$), the Floquet basis and spin basis coincide~\cite{Chen.2020}. If the control fields are switched off precisely at this time, the projective measurements of the spins are equivalent to those of the Floquet basis states. For this reason the spins may be treated as the computational basis, despite the fact that the geometric phase is acquired by the Floquet states. For the experiments performed here, $\Omega_0 / \omega = 1$ setting the magnitude of the phase accumulated $g$, and the loop duration is $T=2\pi/\Omega$ with $\Omega=\Omega_0/10$.

The key feature of Eq.~\ref{eq:mod-ham} is that the entire Hamiltonian (Eq.~\ref{eq:mag-ham}) is modulated by a function with zero time-average, leading to the adiabatic degeneracy of the Floquet states. In any realistic attempt, this condition may be challenged; in our case the presence of stray fields either perturb the spin state energies or couple them, adding terms which do not average to zero over a Floquet period. A fairly general description of stray fields in this system requires only a minor modification of the Hamiltonian (Eq.~\ref{eq:mod-ham}). We consider the case where stray fields are time-independent, since the fields affecting our experiment are stable over several tens of minutes. While the time-dependent case would follow a similar derivation the transformation to the Floquet basis may not be easily performed analytically depending on the specific functional dependence. Equation~\ref{eq:mod-ham} is modified with an extra term,
\begin{equation}
    \hat{H}^{\Delta} = \bm{\Delta} \cdot \bm{\hat{F}},
    \label{eq:delta-ham}
\end{equation}
where we refer to $\bm{\Delta}$ as the detuning from resonance. The $\Delta_z$ component is equivalent to a mismatch between the driving field's carrier frequency and the level-splitting (see Appendix~\ref{s:supp-RF}). Other stray fields or a leaked control field would, in general, correspond to some combination of all three components. In our experiment there is sufficient extinction of the control field, and no other stray fields near resonance, so a $\Delta_z$ term was sufficient to describe the data.

Given that the detuning term in Eq.~\ref{eq:delta-ham} is not modulated by the Floquet envelope, it will result in a new dynamical phase in the Floquet basis despite its lack of time-dependence; this is due to the fact that the Floquet basis transformation is itself time-dependent. If we transform the detuning term into the Floquet basis we obtain (see Appendix~\ref{s:supp-det}),
\begin{multline}
    \hat{H}_{\mathrm{Floq.}}^{\Delta}(t) = 
        \left( 1 - g \right) \bm{\Delta} \cdot \bm{\hat{F}}
        + g \left( \bm{q} \cdot \bm{\Delta} \right)
        \left( \bm{q} \cdot \bm{\hat{F}} \right).
    \label{eq:floq-det}
\end{multline}
The extra detuning breaks the adiabatic degeneracy in the Floquet basis and produces dynamical coupling between states. As in the case of conventional HQC where small perturbations of the energetics break the degeneracy, the presence of any unmodulated term in the Hamiltonian produce similar effects here.
While it may seem that the Floquet driving gives an easy and robust path to degeneracy, the challenges inherent with eliminating terms that do not average to zero are highly analogous to those with maintaining degeneracy in traditional systems~\cite{Zanardi.1999qb}. Furthermore, due to the Floquet basis being a time-dependent mixture of the bare spins, a miscalibration in resonance (which would normally result in a different phase accumulation rate and imperfect population transfer) now results in non-trivial coupling between states. Hence, any undriven term in the lab frame has non-trivial results in the computational basis.

Due to this non-geometric term, we also consider the non-adiabatic generalization~\cite{Anandan.1988} of the holonomy,
\begin{equation}
    \hat{\Gamma}_{A}^{\Delta}(t) = \mathcal{T} \exp\left\{
    i \int_0^T dt \left[ 
    \frac{\partial \bm{q}}{\partial t} \cdot \bm{\hat{A}}(t) - \hat{H}^{\Delta}_{\mathrm{Floq.}}(t) \right]
    \right\},
    \label{eq:nonadi-hol}
\end{equation}
which is represented in time-ordered form, with $\mathcal{T}$ the time-ordering operator. The terms $\partial_t \bm{q} \cdot \bm{\hat{A}}(t)$ and $\hat{H}^{\Delta}_{\mathrm{Floq.}}(t)$ represent the geometric and dynamical contributions to the phase, respectively. While we may write the two sources of phase separately in this way, their impact on the evolution is fundamentally inseparable due to the time-ordering~\cite{Karp.1999}. If we expand this operator, we would find an infinite sum of terms that depend on the nested commutators of the two terms. In certain similar circumstances, there are paths that result in net zero contribution from the dynamical term, which forms the foundation of non-adiabatic HQC~\cite{Sjöqvist.2012}. In our context, since $\bm{\Delta}$ is considered here to be outside of our control, the impact of the detuning given in Eq.~\ref{eq:floq-det} is unlikely to be avoidable. It may be possible through other means, such as dynamical decoupling~\cite{Viola.1999, Khodjasteh.2005, Zanardi.1999, Wu.2022h04, Zhao.2021, Liang.2022}, to suppress the effects of the detuning, but this requires further investigation. We will apply the detuned Hamiltonian (Eq.~\ref{eq:delta-ham}) to the results that follow, demonstrating that even for relatively small detunings, the impact on state evolution is considerable (Fig.~\ref{fig:level-scheme}).

\section{Experimental Results}
\label{s:experiment}

We describe the details of an experiment in which we realize the spin Hamiltonian (Eq.~\ref{eq:mag-ham}) on a specific platform: a non-interacting gas of ultracold \Rb\ atoms, which is similar to that in the proposal of Chen \etal~\cite{Chen.2020}. We explore both of the available stable hyperfine ground states, with total angular momentum quantum numbers $F=1$ and $F=2$; these manifolds have $N=3$ and $N=5$ magnetic sublevels, respectively, with magnetic quantum numbers $\mf = 0, \pm 1, ... \pm F$. We apply a background magnetic field to split the $\mf$ states by $\omega_{\mathrm{Z}} / 2\pi= 1.25\ \mathrm{MHz}$ through the linear Zeeman effect; this field remains fixed for the duration of the experiment. 

After initial laser cooling, forced radio-frequency (RF) evaporation in a magnetic trap, and further evaporation in an optical dipole trap (ODT), we obtain a \Rb\ Bose-Einstein condensate (BEC) of about $10^5$ atoms in the $\ket{F=2, \mf=+2}$ state. To prepare atoms in either of the $F=1$ hyperfine states we use a microwave horn-antenna to couple the $F=1$, $2$ manifolds through a magnetic dipole transition centered at $\approx 6.8\ \mathrm{GHz}$, in addition to RF pulses. For the best population-transfer efficiency and long-term stability, we use 1-ms chirped microwave pulses to effect adiabatic rapid passage (ARP).  In all state preparation sequences, state purity is ensured by intermediate resonant laser pulses that remove atoms remaining in the other hyperfine manifold following the microwave ARP pulse, at the expense of slightly reduced atom numbers.

To produce the spin Hamiltonian (Eq.~\ref{eq:mag-ham}), we use an RF field that couples $\mf$ levels through a magnetic dipole transition on resonance with the $\omega_{\mathrm{Z}}$ splitting. The amplitude, frequency, and phase of this field are periodically modulated with an arbitrary waveform generator (AWG), (details in Appendix~\ref{s:supp-RF}). Each of the loops shown in Table~\ref{tab:loops} is implemented by the appropriate simultaneous variations of $\Theta (t)$ and $\Phi (t)$.

Following each gate operation, we perform state tomography. This is done using Stern-Gerlach (SG) imaging~\cite{Lindon.2023}, in which atoms are spatially separated by a magnetic field gradient after the ODT is turned off, and subsequently absorption imaged to infer the relative spin populations. This constitutes a projective measurement in the spin basis. To gain information about the phase of the spin states, we precede SG measurements with resonant RF pulses of varied phase and pulse areas, which changes the measurement basis, permitting the full tomographic reconstruction of the prepared states. Combined with our ability to initially prepare atoms in each spin basis state, we measure the holonomies in full through a series of informationally complete projections and state-preparation pulses. Given that the absorption imaging process is destructive, each measurement represents the production of a new BEC, a process which takes about $25$ seconds; as such, scans take upwards of 10 minutes depending on how many projections are being taken.

\subsection{Results}
\label{s:results}

\begin{figure*}
    \includegraphics{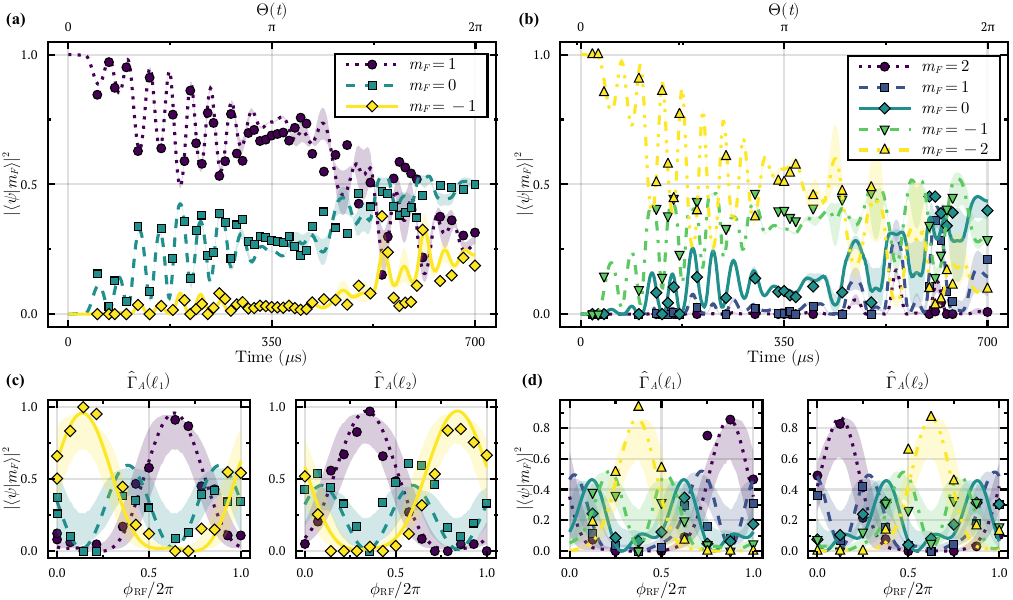}
    \caption{Measured spin populations (points) superimposed onto theory calculations (lines), with $F=1$ (a, c) and $F=2$ (b, d). For all measurements, $\Omega_0 / \omega = 1$, and $\Omega = \Omega_0 / 10$. (a-b) Time evolution throughout the $\hat{\Gamma}_{A}(\ell_1)$ gate, where $\Omega_0 / 2\pi = 14.27$ kHz. (c-d) Comparison of RF-phase scans between $\hat{\Gamma}_{A}(\ell_1)$ (left) and $\hat{\Gamma}_{A}(\ell_2)$ (right), which have identical evolution in the $\hat{F}_z$ basis but clearly differ in the final phase of the state. The field amplitude was $\Omega_0 / 2\pi = 10.64$ kHz and $\Omega_0 / 2\pi = 14.27$ kHz for (c) and (d), respectively. The shaded bands in (a-d) show the effect of the extra detuning term (Eq.~\ref{eq:delta-ham}); for detunings sampled from a Gaussian distribution with mean $\Delta_{\mathrm{fit}}$ (the result of a numerical fit for the detuning $\Delta_z$) and standard deviation $2\pi \times 0.2\ \mathrm{kHz}$, they display the interquartile range of resulting populations. The quadratic Zeeman shift, as described in App.~\ref{s:supp-quadzee}, was included in the numerics (see Sec.~\ref{s:discussion} for discussion).}
    \label{fig:time-series}
\end{figure*}

Using the tomographic measurement techniques described above (\S\ref{s:experiment}) we confirm several aspects of the expected holonomic evolution. The time evolution of the spins during each of the loops was verified by abruptly turning off the control fields, thus interrupting the loop. In changing the time of interruption we map the spin evolution and compare with theory calculations, as shown for $\hat{\Gamma}_{A}(\ell_1)$ in $F=1$ and $F=2$ [Fig.~\ref{fig:time-series}(a-b)]. We see that the evolution over loops results in coupling between states in a way that depends on the loop, as expected with non-Abelian geometric phases.

Additionally, we observe that the phase of the final state also depends on the path taken. We scan a readout RF $\pi / 2$-pulse's phase, comparing two different holonomies, $\hat{\Gamma}_{A}(\ell_1)$ and $\hat{\Gamma}_{A}(\ell_2)$ with $F=1$ and $F=2$ [Fig.~\ref{fig:time-series}(c-d)]. These holonomies produce the same time evolution and final populations in the spin-basis, but the final states differ only in the relative phases between spin components.

Throughout the measurements presented here, background magnetic fields result in a detuning from resonance, as described by Eq.~\ref{eq:delta-ham}. Specifically, the system is susceptible to a $\Delta_z$ component to the detuning, while both other components were negligible. The typical detuning was $\Delta_z / 2\pi \apprle 0.8\ \mathrm{kHz}$, or when compared to the RF coupling strength, $\Delta_z/\Omega_0 \apprle 0.08$. In the absence of magnetic shielding or feedback stabilization of magnetic fields, our ability to control or even detect a detuning of this magnitude is limited; hence, this effect represents a realistic complication in obtaining high-fidelity quantum control (\S\ref{s:fidelity}).

\subsection{Gate Fidelity}
\label{s:fidelity}

For a comprehensive analysis of the holonomic gates, we measured each holonomy in full, focusing on $F=1$ for demonstration. This entailed preparing atoms in each of the spin basis states, applying a holonomy, and performing informationally complete projective measurements, as detailed above. This was done for each of the holonomies in Table~\ref{tab:loops}.

We scanned through a series of state-preparation pulses and measurement pulses for a given gate; the ordering was randomized to prevent any bias coming from a predetermined measurement sequence. The resulting set of projections was then fit for residual detuning, and for the holonomy itself (Appendix~\ref{s:supp-numerics}). We repeated these scans for each holonomy multiple times to account for the randomized detuning which was present in each scan. After each scan we repeated the resonance calibration (Appendix~\ref{s:supp-methods}) in an attempt to detect any drift in the resonance during the measurement. 

\begin{figure}
   \includegraphics{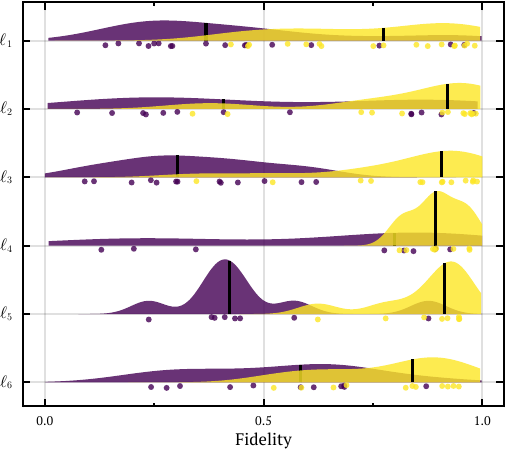}
    \caption{Measured gate fidelities (Eq.~\ref{eq:fidelity}) for each of the loops in Table~\ref{tab:loops}, plotted as density distributions with individual measurements shown scattered beneath. Black vertical bars represent the mean fidelities. A fidelity of one indicates that two transformations are equal. Purple (dark) color show fidelities when detuning is not considered. Yellow (light) color show results when detunings fit to each data set are considered. The quadratic Zeeman shift (App.~\ref{s:supp-quadzee}) was also considered in both measured and predicted holonomies.}
    \label{fig:op-fid}
\end{figure}

From holonomy measurements, we computed the gate fidelities. For a target holonomy $\holonomy{\ell}$, we compute the pure-state fidelity of the measured holonomy $\hat{L}_A (\ell)$ from the inner product,
\begin{equation}
    \label{eq:fidelity}
    \mathcal{F}(\ell) = \frac{\big\vert \mathrm{tr}\left( \hat{L}_A^{\dagger}\hat{\Gamma}_A \right) \big\vert}{2F + 1}.
\end{equation}
If the two matrices are the same then $\mathcal{F} = 1$. Our results are compiled in Fig.~\ref{fig:op-fid}, where we compare our measurements to the target holonomy. Without considering detuning, the fidelities are low, with the average for each loop given in Table~\ref{tab:loops}.

The detuning accurately captures the most significant source of error in our experiment, which we demonstrate by fitting each holonomy measurement scan for detuning $\Delta_z$. Comparing our measured holonomy to a target which includes this detuning in the model (Eq.~\ref{eq:nonadi-hol}) results in much higher fidelities (Fig.~\ref{fig:op-fid}), which are more narrowly distributed, indicating the most significant error mechanism in our apparatus. The average measured fidelities with detuning are also summarized in Table~\ref{tab:loops}. Despite having higher fidelities when detuning is included, the results are still far from unity, with an average of $0.84(7)$. This distribution of fidelities is consistent with our models of shot-to-shot fluctuations in detuning for each measurement of 0.2~kHz; therefore, we find that our infidelity is dominated by detuning fluctuations. The next largest source of error, on the order of about $2\%$, is due to extracting population data from TOF images, which is most significant when the number of atoms in a spin component is low, resulting in an inability to accurately fit for the atomic density distribution.

\subsection{Wilson Loops}
\label{s:wilson-loops}

Another important consideration, in addition to the more practical one of gate fidelity, is to verify the non-Abelian nature of the connection $\bm{\hat{A}}$. To do this, one needs to measure a gauge-invariant manifestation of the non-commutativity between the connection's vector components; in the absence of additional dynamical effects the Wilson loop is appropriate~\cite{Das.2018, Kremer.2019, Sugawa.2021}, which is defined as,
\begin{equation}
    \wilson = \mathrm{tr}\left[ \holonomy{\ell} \right].
    \label{eq:wilson-loop}
\end{equation}
The Wilson loop $\wilson$ is a gauge-invariant measure of the distortions experienced by an eigenbasis through a transformation, $\holonomy{\ell}$. A familiar use of the Wilson loop is in the definition of gate fidelity, as implemented above (\S\ref{s:fidelity}). In this framework, if the measured gate $\hat{L}_A (\ell)$ is the same as the target gate $\holonomy{\ell}$, then the net distortion over the loop $\hat{L}_A^{\dagger} \hat{\Gamma}_A$ is identity, that is, $\wilson = \mathrm{tr}( \identity )$. To show the non-Abelian character of $\bm{\hat{A}}$, one must demonstrate that the Wilson loop is path-dependent, meaning it depends on the ordering of a series of holonomies~\cite{Das.2018}. Due to the cyclic invariance of the trace, three distinct loops must be measured in two different orders, which are non-cyclic permutations of each other. More precisely, for the Wilson loop $\wilson_{ijk}$ with loop order $\ell_i$, $\ell_j$, $\ell_k$, if $\wilson_{ijk} - \wilson_{jik} \neq 0$ then the transformations are path-dependent, and therefore the connection is non-Abelian. A detailed discussion of this may be found in Appendix~\ref{s:supp-trace-comm}. 

\begin{figure}
    \includegraphics{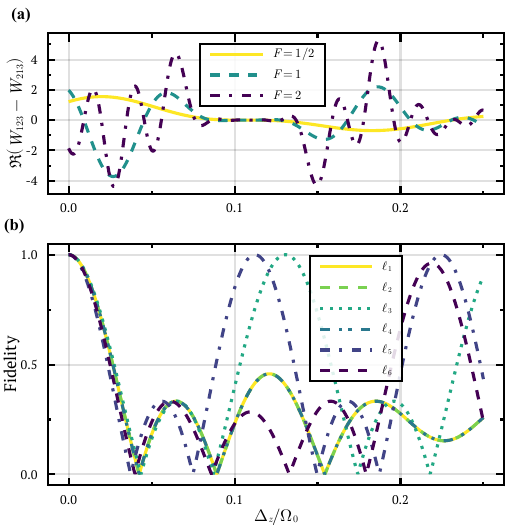}
    \caption{(a) Numerical calculation of difference between Wilson loops for two different path orderings (App.~\ref{s:supp-trace-comm}), which are non-cyclic permutations of each other, as they vary with the detuning $\Delta_z$. $\wilson_{ijk}$ corresponds to consecutive application of loops $\ell_i$, $\ell_j$, and $\ell_k$, respectively. The results vary with the spin manifold $F$ in which the gates are performed. If the Wilson loops depend on the path order, their difference is non-zero, indicating a non-Abelian generator of the transformation. With detuning, there is reduced visibility, and one is unable to separate the impact of non-Abelian dynamical and geometric contributions. (b) Numerical integration of Eq.~\ref{eq:nonadi-hol} with $F=1$ showing dependence of the fidelity for each loop (Table~\ref{tab:loops}) on a $z$-component of the detuning, $\Delta_z$. The fidelities approach zero for relatively small detuning values, but also exhibit periodic revivals.}
    \label{fig:detuning-fidelity}
\end{figure}

Given the significant dynamical contributions arising from detuning in our measured holonomies, as shown in \S\ref{s:fidelity}, we are unable to implement the Wilson loop in this way. The Wilson loop indicates whether the transformations being generated are Abelian or not, but says nothing about the geometric versus dynamical nature of them. Therefore, in a scenario such as this where the dynamical effects are too difficult to isolate, we would be unable to make any strong conclusions on the geometric phase alone. Furthermore, the detuning (Eq.~\ref{eq:delta-ham}) drastically reduces the visibility of the non-Abelian signature $\wilson_{ijk} - \wilson_{jik}$, as its numerical value varies drastically with small values of detuning as shown in Fig.~\ref{fig:detuning-fidelity}(a) , even converging to zero for certain detunings (which would otherwise indicate an Abelian transformation).Therefore, not only is it difficult to accurately measure the path dependence of the Wilson loop, but such a demonstration of path dependence can not be attributed to the geometric nature of the transformation due to the dynamical contributions.

\section{Discussion}
\label{s:discussion}

Our results demonstrate that Floquet-engineering may be used to produce non-Abelian geometric phases in systems which are otherwise non-degenerate. Furthermore, in the context of HQC, the computational basis is conveniently the same as the spin basis due to the stroboscopic nature of the Hamiltonian, without the need for any auxiliary levels or intermediate-state couplings. This technique therefore holds significant potential as an alternative approach to performing holonomic gates, as well as in the generation of interesting artificial gauge fields~\cite{Galitski.2019, Aidelsburger.2018}.

Despite these successes, the Floquet-engineering approach to HQC is not without its limitations. The magnetic field instability present in a typical setup like ours demonstrates a realistic complication in achieving high-fidelity quantum gates, in the form of small miscalibrations in the RF resonances. While Floquet engineering provides a fully degenerate computational basis regardless of the underlying Hamiltonian's energetics, we find that one simply inherits many of the same issues in working with degenerate quantum systems. In fact, it is possible that the degeneracies obtained here are less robust than those in other schemes, such as the dressed-state basis of a tripod scheme~\cite{Zhang.20212cna}.

Our analysis of the effect of detuning also reveals a problem that is more general to all HQC. In the presence of a degeneracy-breaking term like in Eq.~\ref{eq:floq-det}, the traditional form of the holonomy in terms of the Wilczek-Zee non-Abelian phase~\cite{Wilczek.1984, Zee.1988} must be replaced by that in Eq.~\ref{eq:nonadi-hol}~\cite{Anandan.1988, Sjöqvist.2012}. In this representation, modelling or isolating the effects of dynamical noise sources quickly becomes non-trivial, especially in the case of time-dependent noise; this is not just an issue with the Floquet-engineering approach discussed here.

To further characterize the effects of detuning we simulate the gate fidelity for a spin-1 system as it varies with detuning $\Delta_z$ [Fig.~\ref{fig:detuning-fidelity}(b)]. The fidelities fall off quickly with $\Delta_z$, but also exhibit periodic revivals. For the loops $\ell_1$, $\ell_2$, $\ell_4$, and $\ell_6$ the revivals never reach unity for finite $\Delta_z$ due to the dynamical coupling the detuning produces in the Floquet basis. For $\ell_3$, the detuning term is $\propto \hat{F}_z$ in the Floquet basis (Eq.~\ref{eq:floq-det}), and therefore commutes with the geometric portion of the phase. The detuning therefore only changes the relative phase between states, but at certain detunings this net relative phase is $2\pi$ so the fidelity is one. Lastly, for $\ell_5$, the fidelity also approaches one at a particular detuning; however, due the geometric phase not commuting with itself at different times in this loop, a closed-form solution of the holonomy $\hat{\Gamma}_A^{\Delta} (\ell_5)$ (Eq.~\ref{eq:nonadi-hol}) is unknown. As such, this was verified numerically.

Given the effects of the detuning it is essential in future implementations to reduce the relative detunings. It should be noted, however, that simply increasing the Rabi-frequency $\Omega_0$ would be insufficient on its own: while this would permit faster gate operations (scaling the Floquet drive $\omega$ and gate frequency $\Omega = 2\pi/T$ accordingly), the ability to detect detuning as outlined in Appendix~\ref{s:supp-methods} decreases with $\Omega_0$. Therefore, calibrating the gates in this way would result in similar detunings relative to the Rabi frequency. To avoid this issue, one also needs to adopt a different detection technique. There are many magnetometry techniques that could be readily implemented~\cite{Budker.2007}, but they may require additional hardware outside of the gate control-scheme itself (such as an external laser for Faraday magnetometry). Therefore, for a more scalable quantum computing setup one should calibrate the resonance with the gate-control scheme itself, as in our experiment. For instance, measuring populations after an RF-pulse with pulse area $\Omega_0 t=n\pi$ with $n$ an odd integer. For large values of $n$ the detuning sensitivity increases. Ideally, one would apply pulses with much longer durations than their gates. This change, coupled with the more obvious additions of magnetic shielding and/or stabilization should be sufficient to realize Floquet-engineered gates with competitive fidelities, in addition to a proper characterization of the Wilson loop as outlined in \S\ref{s:wilson-loops}. For instance, with the magnetic field stability reported in similar ultracold atom systems~\cite{Sugawa.2021}, we expect the fidelities of the gates shown here would exceed 0.99. The gate fidelity expected in other quantum computing platforms would require detailed analysis, but would follow from the detuning calculations provided here (Eq.~\ref{eq:nonadi-hol}).

It is also important in principle to consider the effects of the quadratic Zeeman shift, which is a second-order correction to the spin-state energies proportional to the linear Zeeman splitting, $\omega_{\mathrm{Z}}$ (Appendix~\ref{s:supp-quadzee}). Like the detuning, this term is unmodulated by the Floquet envelope and therefore leads to dynamical degeneracy-breaking terms in the Floquet Hamiltonian; however, unlike the detuning, the magnitude of this additional shift is approximately constant across all measurements, and can therefore be treated as a systematic effect rather than a transient miscalibration (as in the case of detuning). This effect was included in the analyses presented here, however, since this term only appears in higher-spin systems and thus does not generalize to qubits, in the generic context of HQC it is irrelevant. The generalized detuning described in Eq.~\ref{eq:delta-ham} and App.~\ref{s:supp-det} are sufficient to describe time-independent calibration errors that may occur in traditional qubit implementations.

Despite these limitations, Floquet-engineered gates may still have an important place in the quantum-control engineer's toolbox. There are many examples of experimental systems with excellent isolation of background magnetic fields and qubit-resonances~\cite{Sugawa.2021, Borkowski.2023, Xu.2019}. This is typically a necessary step in precise quantum control regardless of what gate architecture is used. It is unlikely that any gates are tolerant to all faults (even in the case of topological gates~\cite{Lahtinen.2017, Nayak.2008}), and so eliminating their shortcomings will always be required. Gates that exhibit some level of fault-tolerance are therefore welcome in reducing the complexities of quantum computers. In the case of the Floquet-engineered gate demonstrated here, more work must be done to overcome coherent and decoherent noise sources~\cite{Li.2021e63, Liu.2023gd, Kang.202272u, Xu.2012, Oreshkov.2009, Kang.2023}, in order to achieve proposed performance that is tolerant to high-frequency fluctuations, as is typical for geometric gates~\cite{Chen.2020, Wang.2021}.

Compared to alternate approaches to HQC~\cite{Zhang.20212cna}, Floquet engineering offers the significant advantage of choosing a convenient computational basis without the need for auxiliary levels or intermediate-state couplings. While the degeneracies and phases are best represented in the Floquet basis, its stroboscopic coincidence with the bare states (spin basis in this case) provides experimental simplicity in state preparation and measurement. This scheme has already been generalized to multi-qubit control in Rydberg atoms~\cite{Wang.2021}, and could be similarly extended to other popular quantum computing platforms which host sufficient control mechanisms.

\section{Conclusion}
\label{s:conclusion}

We demonstrated and characterized an approach to HQC in an ultracold ensemble of \Rb. Through periodic modulations of a control Hamiltonian, the resulting Floquet-engineered system behaves as though it is fully degenerate regardless of the underlying level structure. Using adiabatic evolution of control parameters, we generated non-Abelian geometric phases for the purpose of universal single-qubit quantum gates. While our demonstration used the entire $F=1$ or $F=2$ ground state manifolds containing three and five spin states respectively, the control Hamiltonian generates transformations in $SU(2)$, and is therefore readily applicable to any two-level system. Our demonstration was limited to the context of single-qubit gates, but the approach, including the detuning analysis, may be readily generalized to two-qubit gates in a similar manner~\cite{Wang.2021}.

Further study could also illuminate how the scheme might generalize to arbitrary $SU(2F+1)$ state transformations: additional coupling fields and a non-linear splitting between levels, such as through the quadratic Zeeman effect, could yield connections with such a symmetry. Such schemes are of interest to the control of qudits, or in the generation of new artificial gauge fields.

In our experimental demonstration, the existence of ambient magnetic field fluctuations revealed the limits of the approach. While Floquet engineering quite easily produces degeneracies in systems which would otherwise not support them, the technique inherits many of the same difficulties that come with maintaining them. This resulted in limited gate fidelity. Despite this, these gates offer the freedom to encode information on any readily applicable basis, without the need for auxiliary levels, removing a level of difficulty in state readout and preparation. Furthermore, the degree of fault-tolerance from sources other than static-detuning has not been investigated in great detail. Clearly, the Floquet-engineering approach deserves ongoing attention.

\begin{acknowledgments}

{This work was supported by the University of Alberta; the Natural Sciences and Engineering Research Council (NSERC), Canada (Grants No. RGPIN-2021-02884 and No. CREATE-495446-17);  the Alberta government's Quantum Major Innovation Fund; Alberta Innovates; the Canada Foundation for Innovation, and the Canada Research Chairs (CRC) Program.
We gratefully acknowledge that this work was performed on Treaty 6 territory, and as researchers at the University of Alberta, we respect the histories, languages, and cultures of First Nations, M\'etis, Inuit, and all First Peoples of Canada, whose presence continues to enrich our vibrant community.}
\end{acknowledgments}

% Supplemental
%\clearpage

\appendix
\renewcommand{\thefigure}{A\arabic{figure}}
\setcounter{figure}{0}
\label{s:supplemental}

\section{RF Dressing}
\label{s:supp-RF}

\begin{figure}[h]
    \includegraphics{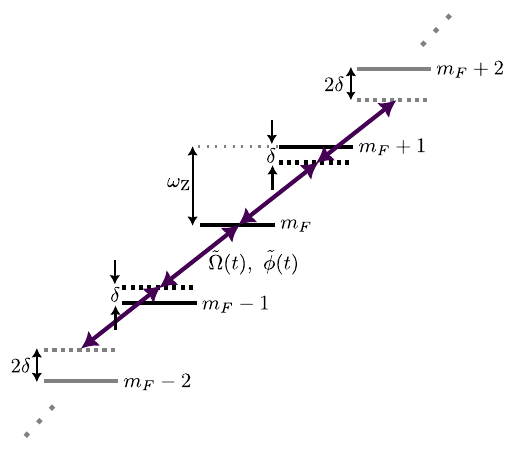}
    \caption{Spin manifold $F$ with $N= 2F + 1$ magnetic sublevels, denoted $m_F$, that are separated in energy by $\omega_{\mathrm{Z}}$ through the Zeeman effect. Adjacent levels are coupled by an RF magnetic field (Eq.~\ref{eq:mag-field}) with Rabi frequency $\tilde{\Omega}(t)$ and phase $\tilde{\phi}(t)$, both of which are modulated in time to trace out loops in parameter space (Eq.~\ref{eq:amp-mod} and Eq.~\ref{eq:phase-mod}). The frequency of the driving RF field $\omega_{\mathrm{RF}}$ is modulated close to resonance with the level splitting, $\delta \approx \omega_{\mathrm{Z}} - \omega_{\mathrm{RF}}$ (Eq.~\ref{eq:modulated-detuning}).}
    \label{fig:rf-levels}
\end{figure}

Here we detail the derivation of the Hamiltonian (Eq.~\ref{eq:mag-ham}).  We begin with $N = 2F +1$ spin levels subjected to a static magnetic field along the $z$-direction that splits, through the Zeeman effect, adjacent levels by $\omega_{\rm Z}$, as shown in Fig.~\ref{fig:rf-levels}. These spins are subject to an RF magnetic field in the $x$-direction,
\begin{equation}
    \label{eq:mag-field}
    B_{\rm RF}(t) = \tilde{B}(t) \sin\left[ \omega_{\mathrm{RF}} t + \tilde{\phi}(t)\right]
\end{equation}
that couples adjacent spin levels, where $\omega_{\mathrm{RF}}$ is the RF-carrier frequency, and $\tilde{\phi}(t)$ the phase. In the dressed-states basis, the Hamiltonian in manifold $F$ (in units of $\hbar=1$) in the lab-frame, is
\begin{equation}
    \label{eq:supp-lab-ham}
    \hat{H}_{\mathrm{RF}} = \tilde{\Omega}(t) \sin \left[ \omega_{\mathrm{RF}} t + \tilde{\phi}(t) \right] \hat{F}_x + \omega_{\mathrm{Z}} \hat{F}_z,
\end{equation}
where $\tilde{\Omega}(t) = \langle m_F | g_F \mu_{\mathrm{B}} \tilde{B}(t) | m_F \pm 1\rangle$ is the Rabi frequency associated with the coupling matrix element for any adjacent levels, with $g_F$ the hyperfine $g$-factor, and $\mu_{\mathrm{B}}$ the Bohr magneton. Both the phase of the driving field $\tilde{\phi}(t)$ and the amplitude $\tilde{\Omega}(t)$ are modulated in time.

Next, we transform this Hamiltonian (Eq.~\ref{eq:supp-lab-ham}) into the frame rotating with the RF carrier frequency $\omega_{\mathrm{RF}}$ (with frequency modulation). The Hamiltonian transforms according to,
\begin{equation}
    \label{eq:supp-basis-transform}
    \hat{H}_{\mathrm{R}} = \hat{U} \hat{H}_{\mathrm{RF}} \hat{U}^{\dagger} -i \hat{U}^{\dagger} \partial_t \hat{U},
\end{equation}
with the unitary $\hat{U}(t) = \exp\left\{ -i \left[ \omega_{\mathrm{RF}} t + \gamma (t) \right] \hat{F}_z \right\}$. Here we have transformed with an arbitrary phase of rotation $\gamma (t)$, which will be determined by the frequency modulations computed later. We may compute the transformation (Eq.~\ref{eq:supp-basis-transform}) by rewriting in terms of the raising and lowering operators, $\hat{F}_x = \left( \hat{F}_+ + \hat{F}_- \right) / 2$, and using,
\begin{align}
    \hat{U} \hat{F}_z \hat{U}^{\dagger} &= \hat{F}_z, \\
    \hat{U} \hat{F}_{\pm} \hat{U}^{\dagger} &= e^{\pm i \left( \omega_{\mathrm{RF}} t + \gamma \right)} \hat{F}_{\pm}.
\end{align}
We find that
\begin{align}
    \hat{H}_{\mathrm{R}} = \frac{i \tilde{\Omega}}{4} &\Bigg\{ 
        \Big[ e^{i \left( \gamma - \tilde{\phi} \right)} - e^{i \left( 2\omega_{\mathrm{RF}} t + \tilde{\phi} + \gamma \right)} \Big] \hat{F}_+\nonumber\\
        & + \left[ e^{-i \left( 2\omega_{\mathrm{RF}} t + \tilde{\phi} + \gamma \right)} - e^{-i \left( \gamma - \tilde{\phi} \right)}\right] \hat{F}_- \Bigg\}\nonumber\\
        &+ \left[ \omega_{\mathrm{Z}} - \left( \omega_{\mathrm{RF}} + \partial_t \gamma \right) \right] \hat{F}_z.
\end{align}
Since the RF carrier frequency $\omega_{\mathrm{RF}} / 2\pi = 1.25\ \mathrm{MHz}$ is large compared to the Rabi frequency $\vert\tilde{\Omega}\vert / 2\pi \approx 10\ \mathrm{kHz}$ then we may apply the rotating wave approximation by ignoring the terms which rotate at $2\omega_{\mathrm{RF}}$. Simplifying what remains in terms of the spin operators we have,
\begin{align}
    \hat{H}_{\mathrm{R}}(t) = \frac{\tilde{\Omega}(t)}{2} \Big\{ &
        \sin\left[ \tilde{\phi} (t) - \gamma (t) \right] \hat{F}_x\nonumber\\
       & - \cos\left[ \tilde{\phi} (t) - \gamma (t) \right] \hat{F}_y
        \Big\}
        + \delta (t) \hat{F}_z, \\
    \delta (t) =& \quad \omega_{\mathrm{Z}} - \omega_{\mathrm{RF}} - \partial_t \gamma (t). \label{eq:modulated-detuning}
\end{align}
Looking at the form of the target Hamiltonian (Eq.~\ref{eq:mag-ham}) we determine the form of the required amplitude and phase modulations, $\tilde{\Omega}(t)$, and $\tilde{\phi}(t)$ respectively. First, by inspection, we can see that,
\begin{equation}
    \label{eq:amp-mod}
    \tilde{\Omega}(t) = 2 \Omega_0 \sin\left[ \Theta (t) \right] \cos \omega t,
\end{equation}
with $\Theta (t)$ the polar angle for paths in parameter space. Looking then at the $\hat{F}_z$ term, we require that,
\begin{equation}
    \delta (t) = \Omega_0 \cos\left[ \Theta(t) \right] \cos \omega t.
\end{equation}
Ideally, the driving RF field is resonant with the level splitting, $\omega_{\mathrm{RF}} = \omega_{\mathrm{Z}}$; when this condition is not properly met, an extra detuning term (Eq.~\ref{eq:delta-ham}) must be considered (see \S\ref{s:results} and \S\ref{s:discussion}). Assuming we meet this resonance condition, we obtain,
\begin{equation}
    \partial_t \gamma (t) = -\Omega_0 \cos\left[ \Theta (t) \right] \cos \omega t
\end{equation}
so then,
\begin{equation}
    \gamma(t) = - \Omega_0 \int_0^t dt^{\prime} \cos\left[ \Theta (t^{\prime}) \right] \cos \omega t^{\prime}.
\end{equation}
Altogether, we obtain for the phase modulation term,
\begin{equation}
    \label{eq:phase-mod}
    \tilde{\phi}(t) = \Phi(t) + \frac{\pi}{2} - \Omega_0 \int_0^t dt^{\prime} \cos\left[ \Theta (t^{\prime}) \right] \cos \omega t^{\prime},
\end{equation}
where $\Phi(t)$ is the azimuthal angle for paths in parameter space, and the $\pi / 2$ term is a phase shift necessary to obtain the desired phase reference in the Hamiltonian (Eq.~\ref{eq:mag-ham}). From this we can see how one generates the required RF waveforms in the lab-frame, complete with phase and amplitude modulations, $\tilde{\phi}(t)$, and $\tilde{\Omega}(t)$, in terms of an arbitrary path $\ell$ parameterized by $\Theta (t)$, and $\Phi (t)$.

\section{Floquet Hamiltonian}
\label{s:supp-floquet}

In order to derive the Floquet Hamiltonian Eq. \ref{eq:floq-ham} from the rotating basis Hamiltonian $\hat{H} = \hat{V}(t)\cos \omega t $ (as laid out in Ref. \cite{Novičenko.2019}), we first
define a micromotion operator
\begin{align}
  \label{eq:micromotion1}
  \hat{U} = \exp\left[ -i \hat{V}(t) \sin(\omega t)/\omega \right],
\end{align}
where the $\sin(\omega t)/\omega$ is chosen such that it averages to zero, and its derivative is equal to the Floquet drive $\cos(\omega t)$. Every $t = 2n\pi/\omega$ we have that $\hat{U}(t) = {\identity}$ so that the basis induced by $\hat{U}$ coincides with the original basis, meaning they are {\sl stroboscopically} equivalent. Transforming to the basis induced by Eq. \ref{eq:micromotion1} gives
\begin{align}
  i \partial_{t} \ket{\phi} &= \hat{H} \ket{\phi}, \nonumber \\
  i \partial_{t} \left(\hat{U} \hat{U}^\dagger  \ket{\phi} \right)&= \hat{H} \hat{U} \hat{U}^\dagger \ket{\phi},  \nonumber \\
  \hat{U}^\dagger i \partial_{t} \left(\hat{U} \hat{U}^\dagger  \ket{\phi} \right)&= \hat{U}^\dagger \hat{H} \hat{U} \hat{U}^\dagger \ket{\phi}. \nonumber 
\end{align}
Defining $\ket{\psi} = \hat{U}^\dagger \ket{\phi}$ and applying the chain rule to the $\partial_t$ term, 
\begin{align}
  i\partial_t\ket{\psi} = \left( \hat{U}^\dagger\hat{H} \hat{U} - i \hat{U}^\dagger {\partial_t\hat{U} \over \partial t} \right)\ket{\psi},
\end{align}
such that the Hamiltonian in the basis of the micromotion operator Eq. \ref{eq:micromotion1} becomes
\begin{align}
  \hat{H}_U(t) &= \hat{U}^\dagger\left( \hat{H} - i\partial_t \right) \hat{U}, \nonumber  \\
         &=   \hat{H} - i \hat{U}^\dagger {\partial \hat{U} \over \partial t}, 
\end{align}
Where we have used the relationship that $\hat{H} \propto \hat{V}$, such that $[\hat{H}, \hat{U}] = 0$. We can then use a version of the Baker-Hausdorff
lemma for differential operators
\begin{align}
  -e^{\hat{X}} \,{\partial  e^{-\hat{X}} \over \partial t}  
  =  {\partial \hat{X} \over \partial t} 
  &+ {1 \over 2!}\left[\hat{X}, {\partial \hat{X} \over \partial t}\right] \nonumber \\
  &+ {1 \over 3!} \left[\hat{X} , \left[\hat{X}, {\partial \hat{X} \over \partial t}\right] \right] + ...~~,
\end{align}
\begin{widetext}
and define $c = \sin(\omega t)/\omega$ to write
\begin{align}
  \hat{H}_U(t) &=  \hat{V}\cos(\omega t) - i \hat{U}^\dagger {\partial_t\hat{U} \over \partial t} \nonumber \\
         &= \hat{V}\cos(\omega t)
           + i \left(  i {\partial \hat{V} c\over \partial t} 
           + {i^2 c \over 2!}\left[\hat{V} , {\partial \hat{V} c\over \partial t} \right]
           + {i^3 c^2 \over 3!} \left[\hat{V} , \left[\hat{V}, {\partial \hat{V} c\over \partial t}\right]\right] + ... \right) \nonumber \\
         &= \hat{V}\cos(\omega t) - \hat{V}\cos(\omega t)
           +i \left(  i c \, {\partial_t \hat{V}}
           - {(ic)^2 \over 2! }\left[\hat{V}, {\partial_t \hat{V}}  \right]
           + {(ic)^3 \over 3!}\left[\hat{V}, \left[\hat{V}, {\partial_t \hat{V}} \right]  \right] + ...\right) \nonumber \\
         &= i \left(  i c \, {\partial_t \hat{V}}
           + {(ic)^2 \over 2! }\left[\hat{V}, {\partial_t \hat{V}}  \right]
           + {(ic)^3 \over 3!}\left[\hat{V}, \left[\hat{V}, {\partial_t \hat{V}} \right]  \right] + ...\right), 
\end{align}
where we have used that $[\hat{V}, \hat{V}\, \partial_t c] = 0$ to only consider the derivatives with respect to $\hat{V}$ for the terms involving commutators. 
We now notice that the iterated commutators between $\hat{V}$ and $\partial_t \hat{V}$ form a recurrence relation
\begin{align}
  [\hat{V}, \partial_t \hat{V}]
  &= i \Omega_0^2 \, (\bm{q}\times \partial_t \bm{q})\cdot \bm{\hat{F}}, \\
  \left[\hat{V} , [\hat{V}, \partial_t \hat{V}] \right]
  &= \Omega_0^3 \, \partial_t \bm{q} \cdot \bm{\hat{F}},\\
  \left[ \hat{V}, \left[\hat{V} , [\hat{V}, \partial_t \hat{V}] \right] \right]
  &= \Omega_0^2 \,  [\hat{V}, \partial_t \hat{V}],
\end{align}
so that
\begin{align}
  \hat{H}_U(t) &= i\left[ (\partial_t \bm{q})\cdot\bm{\hat{F}} \, {i \Omega_0 c \over 1!}
           + i(\bm{q}\times \partial_t \bm{q})\cdot \bm{\hat{F}} \, {(i \Omega_0 c)^2 \over 2!} 
           + \partial_t \bm{q} \cdot \bm{\hat{F}} \, {(i \Omega_0 c)^3 \over 3!} + ... \right]\nonumber \\
         &= i \left[
           (\partial_t \bm{q}) \cdot \bm{\hat{F}} \, \sum_{n=1}^\infty {(ic\Omega_0)^{2n-1} \over (2n-1)!}
           + i(\bm{q}\times \partial_t \bm{q})\cdot \bm{\hat{F}} \, \sum_{n=1}^\infty {(ic\Omega_0)^{2n} \over (2n)!}
           \right]\nonumber \\
         &= - (\partial_t \bm{q}) \cdot \bm{\hat{F}} ~  \sin\left(\Omega_0 {\sin \omega t \over \omega}  \right)
           - (\bm{q}\times \partial_t \bm{q})\cdot \bm{\hat{F}} ~ \left[ \cos\left(\Omega_0 {\sin \omega t \over \omega}  \right) -1  \right].
\end{align}
\end{widetext}

Now, we expand $\hat{H}_U(t)$ in a {\sl quasistatic} Fourier series in $\omega t$, essentially assuming that $\hat{V}(t)$ is approximately
constant over a $2\pi/\omega$ period:
\begin{align}
  f(t) &= \sum_{n=-\infty}^\infty f^{(n)}(t)e^{i n \omega t},\\
  f^{(n)}(t) &= {\omega \over 2\pi} \int_{0}^{2\pi/\omega} \mathrm{d}t' f(t + t') e^{-i n\omega t},
\end{align}
so that the Schr\"odinger equation 
\begin{align*}
 i \partial_t \ket{\psi(t)} = \hat{H}_{U}(t) \ket{\psi(t)}
\end{align*}
transforms into
\begin{align*}
 ( i\partial_{t} - n \omega t) \ket{\psi^{(n)}(t)} = \sum_{nm}\hat{W}^{(n-m)} \ket{\psi^{(m)}(t)},
\end{align*}
or
\begin{align}
 i\partial_{t}  \ket{\psi^{(n)}(t)} = \sum_{nm}\left(n \omega \delta_{n,m} + \hat{H}^{(n-m)}_{U} \right)\ket{\psi^{(m)}(t)}.
\end{align}
We call
\begin{align}
  \hat{H}_{\rm Floq}^{(n-m)}(t) = n \, \omega \delta_{n,m} + \hat{H}^{(n-m)}_{U}(t), 
\end{align}
the Floquet Hamiltonian, where the fast oscillations in the original Hamiltonian have been factored out and replaced with
a new set of quantum numbers $(n)$ which are analogous to band indices from materials with periodic spatial structure.

To perform the quasistatic Fourier transform on $\hat{H}_{\rm Floq}^{(n)}$, we use the identities
\begin{align}
  {1 \over 2\pi} \int_{0}^{2\pi}\mathrm{d}\theta \, e^{i n \theta}\sin\left(a \sin(\theta) \right) &= i\left( {1 - (-1)^n \over 2} \right) J_{n}(a), \\
  {1 \over 2\pi} \int_{0}^{2\pi}\mathrm{d}\theta \, e^{i n \theta}\cos\left(a \sin(\theta) \right) &= \left( {1 - (-1)^{n+1} \over 2} \right) J_{n}(|a|).
\end{align}
Now, noting that since $|J_{n}(x)| \leq 0$, the matrix elements of $\hat{H}_{U}^{(n)}$ are strictly bounded,  
\begin{align}
\langle \alpha| \hat{H}_{U}^{(n)} | \beta \rangle \leq F \, |\partial_t \bm{q}|,
\end{align}
where $F$ is the spin quantum number corresponding to $\bm{\hat{F}}$. In the slow limit where $F \,  |\partial_t \bm{q}| \ll \omega$, we may then neglect 
couplings between different Floquet bands, since the energetic gap between different bands will be very large relative to the off-diagonal terms. With 
transitions between states in different Floquet bands suppressed,  we may focus on the evolution of a state solely {\sl within} a given band. 
Therefore, we choose to restrict all attention to the $n=0$ Floquet band, where the Floquet Hamiltonian becomes: 
\begin{align}
  \hat{H}_{\rm Floq}(t) &= {\omega \over 2\pi} \int_{0}^{2\pi/\omega} \mathrm{d}t' \, H_{U}(t + t')\nonumber\\
               &= \left[1 - J_0\left({\Omega_0 \over \omega}  \right)  \right] \left[\bm{q}(t)\times \partial_t \bm{q}(t) \right] \cdot \bm{\hat{F}} \nonumber\\
               &= {\partial \bm{q} \over \partial t} \cdot \left\{
                 \left[1 - J_0\left({\Omega_0 \over \omega}  \right)  \right] \bm{\hat{F}} \times \bm{q}(t)
                 \right\}\nonumber\\
               &= {\partial \bm{q} \over \partial t} \cdot \bm{\hat{A}}(t),
\end{align}
and we have defined
\begin{align}
\bm{\hat{A}}(t) = \left[1 - J_0\left({\Omega_0 \over \omega} \right)\right]\bm{\hat{F}}\times \bm{q}(t),
\end{align} 
the non-Abelian {\sl connection}. This agrees with the derivation in Ref. \cite{Novičenko.2019} for the special case where $\bm{q}$ is of constant unit length.

In the Floquet basis, the time-ordered evolution operator generated by $\hat{H}_\mathrm{Floq.}$ is given by:
\begin{align}
    \hat{U}(t-t_0) 
    &= \mathcal{T} \exp\left(-i\int_{t_0}^t \mathrm{d}t' \, \hat{H}_{\mathrm{Floq.}}(t') \right) \nonumber \\
    &= \mathcal{T} \exp\left(-i\int_{t_0}^t \mathrm{d}t' \, {\partial \bm{q}(t') \over \partial t'} \cdot \bm{\hat{A}}(t') \right).
\end{align}
Now, we may perform a change of variables to replace the explicit time-ordering and an integral over $\mathrm{d}t'$ to an integral over $\mathrm{d}\bm{q}$ itself with {\sl path-ordering},
\begin{align}
    \hat{U}(\ell) 
    &= \mathcal{P} \exp\left(-i\int_\ell \, \mathrm{d}\bm{q} \cdot \bm{\hat{A}} \right),
\end{align}
where $\ell$ is the {\sl path} traced out by $\bm{q}$ from times $t_0$ to $t$. In this way, due to the special structure of $\hat{H}_{\mathrm{Floq.}}$, the evolution of the system may be interpreted fully geometrically, depending only on the path traced out by $\bm{q}$ so long as the system is in the adiabatic limit. 

For a {\sl closed} path $\ell$, $U(\ell)$ is known as a {\sl holonomy} which encodes geometric and topological information about the Hilbert space and the group generated by $\bm{\hat{A}}$. We typically denote holonomies with the symbol $\Gamma_A$,
\begin{align}
    \hat{\Gamma}_A(\ell) 
    &= \mathcal{P} \exp\left(-i\oint_\ell \, \mathrm{d}\bm{q} \cdot \bm{\hat{A}} \right)~~.
\end{align}
The {\sl trace} of a holonomy is a gauge and basis invariant quantity known as a {\sl Wilson loop}:
\begin{align}
    \mathcal{W}(\ell) = \mathrm{tr}\left[ \hat{\Gamma}_A(\ell)\right]~~.
\end{align}

\section{Trace Commutator}
\label{s:supp-trace-comm}

As demonstrated in Ref. \cite{Das.2018}, showing that the evolution operators for two distinct loops do not commute is insufficient to prove that a system has truly path-dependent evolution. The authors describe situations where basis-dependent effects can
cause two evolution operators from an Abelian gauge theory to fail to commute. They then argue that to properly determine that an atomic gas
hosts a non-Abelian geometric phase, one should measure basis-independent quantities, such as the difference between two Wilson loops with different orders:
\begin{align} \label{eq:tr_comm}
  I &= \mathrm{tr}\Big( \hat{\Gamma}_A(p_3) \left[\hat{\Gamma}_A(p_2), \hat{\Gamma}_A(p_1) \right] \Big)\\
    &=  \wilson_{123} - \wilson_{213},
\end{align}
for three independent loops $p_1$, $p_2$, $p_3$. If $I \neq 0$, then the connection $\hat{\Gamma}_A$ is truly non-Abelian 
regardless of any gauge or basis dependent effects. One choice of the paths $p_i$ which are useful to work with are the set of unit-radius great circles:
\begin{align}
  p_1 &= \{\Phi(t) = -\pi/2 ,\; \Theta(t) = \Omega t\}, \\
  p_2 &= \{\Phi(t) = 0 ,\;\, ~~~~~\Theta(t) = \Omega t\},  \\
  p_3 &=  \{\Phi(t) = \Omega t ,\;\, ~~~\Theta(t) = \pi/2 \},
\end{align}
which are chosen such that for $\bm{q}$ along these paths, $\bm{\hat{F}} \times \bm{q}$ is constant along the $\bm{q}$ direction. Note that this set of loops produce nearly equivalent transformations as the loops considered in the experiment (Table~\ref{tab:loops}), with $\ell_1$ identical to $p_2$, $\ell_2$ being the reverse of $p_1$, and $\ell_3$ identical to $p_3$. The path-ordered evolution operator in Floquet basis can be written as:
\begin{align}
  \hat{\Gamma}_A(p_i) &= \mathcal{P} \exp\left( -i \,\oint_{p_i}\mathrm{d}\bm{q}\cdot \, \bm{\hat{A}} \right) \nonumber \\
  &= \mathcal{P} \exp\left[ -i \,g \, \oint_{p_i}\mathrm{d}\bm{q}\cdot \, \left(\bm{\hat{F}} \times \bm{q}\right) \right],
\end{align}
but if $\bm{\hat{F}} \times \bm{q}$ is constant along $p_i$, then the path ordering becomes unnecessary and the evolution operator becomes:
\begin{align}
  \hat{\Gamma}_A(p_i) &= \exp\left[ -i \,g \, \oint_{p_i}\mathrm{d}\bm{q}\cdot \, \left(\bm{\hat{F}} \times \bm{q}\right) \right] \nonumber \\
  &= \exp\left( -i \,2\pi \, g \, \bm{\hat{F}}_i \right) ~~.
\end{align}
Via Stokes' theorem, this can be interpreted as a non-Abelian {\sl flux} of $2\pi \, g \, \bm{\hat{F}}_i$ through the loop $p_i$.

\begin{widetext} 
Noticing that the operators $\hat{\Gamma}_A(p_i)$ are $SU(2)$ rotation operators, we decompose them in terms of Euler angles in the $Z-Y-Z$ convention:
\begin{align}
  \hat{\Gamma}_A(p_1) &= e^{i {\pi \over 2} \hat{F}_3} e^{-i \, g \, \hat{F}_2} e^{-i {\pi \over 2} \hat{F}_3}
  = \hat{\mathcal{R}}\left( -{\pi \over 2}, 2\pi \, g , {\pi \over 2} \right), \\
  \hat{\Gamma}_A(p_2) &= e^{-i 0 \hat{F}_3} e^{-i \, 2\pi g \, \hat{F}_2} e^{-i 0 \hat{F}_3}
  = \hat{\mathcal{R}}\left(0,  \, 2\pi \, g \,, 0 \right), \\
  \hat{\Gamma}_A(p_3) &= e^{-i \, 2\pi \, g \, \hat{F}_3} e^{-i 0 \hat{F}_2} e^{-i 0 \hat{F}_3}
  = \hat{\mathcal{R}}\left(2\pi \, g \,, 0, 0 \right).
\end{align}
The matrix elements of a rotation operator can be found in terms of Wigner's $d$-matrix \cite{sakurai2014modern} as:
\begin{align}
  \braket{F,m \Big| \hat{\mathcal{R}}(\alpha, \beta, \gamma) \Big|F, m'} = e^{-i\alpha m} d^{F}_{m,m'}(\beta) e^{-i\gamma m\prime}, 
\end{align}
where $F$ is the total spin quantum number corresponding to $\bm{\hat{F}}$, and:
\begin{align}
  d^{F}_{m,m'}(\beta)
  &= \braket{F,m \Big| e^{-i \beta \hat{F}_2} \Big| F, m'} \nonumber \\
  &= \sqrt{(F+m')!\, (F-m')!\, (F+m)!\, (F-m)!}\sum_{k=k_{\mathrm{min}}}^{k_{\mathrm{max}}}
    \left[\frac{(-1)^{m'-m+k} \left(\cos\frac{\beta}{2}\right)^{2F+m-m'-2k}\left(\sin\frac{\beta}{2}\right)^{m'-m+2k} }{(F+m-k)!\, (F-m'-k)!\, (m'-m+k)!\, k!} \right],
\end{align}

and $k_{\mathrm{min}} = \mathrm{max}(0, m-m')$, $k_{\mathrm{max}} = \mathrm{min}(F+m, F-m)$. These matrices have the properties
\begin{align}
  d^{F}_{m,m'}(0) &= \delta_{m,m'}, \\
  d^{F}_{m,m'}(\beta) &= (-1)^{m-m'}d^F_{m',m} = d^{F}_{-m',-m},
\end{align}
and are manifestly real in this basis. Thus, using the $d$-matrices, we can write
\begin{align}
  \mathrm{tr}\Big( \hat{\Gamma}_A(p_3) \hat{\Gamma}_A(p_2) \hat{\Gamma}_A(p_1) \Big)
  &= \sum_{m}\braket{F,m|\hat{\mathcal{R}}\left(2\pi, g, 0, 0 \right) \hat{\mathcal{R}}\left(0,  2\pi, g, 0 \right) \hat{\mathcal{R}}\left( -{\pi \over 2}, 2\pi, g, {\pi \over 2} \right) | F, m} \nonumber\\
  &= \sum_{m m' m''} \braket{F,m|\hat{\mathcal{R}}\left(2\pi, g, 0, 0 \right)| F, m'}
    \braket{F,m'| \hat{\mathcal{R}}\left(0,  2\pi, g, 0 \right)|F,m''} \nonumber \\
    &~~~~~~~~~~~~~~~~~
    \times \braket{F,m'' | \hat{\mathcal{R}} \left( -{\pi \over 2}, 2\pi, g, {\pi \over 2} \right) | F, m} \nonumber\\
  &= \sum_{m m' m''} e^{-2\pi i\,m\, g}d^{F}_{m,m'}(0) \, d^{F}_{m',m''}(2\pi \, g) \, e^{i m'' \pi /2}d^{F}_{m'',m}(2\pi \, g) \, e^{-i m \pi/2} \nonumber \\
  &=  \sum_{m, m'} e^{i {\pi \over 2}(m'-m)}  e^{-2\pi i\,m\, g}  d^{F}_{m,m'}(2\pi \, g) \, d^{F}_{m',m}(2\pi \, g)~~,
\end{align}
where the sums of $m$ and $m'$ run from $-F$ to $F$. Similarly
\begin{align}
    \mathrm{tr}\Big( \hat{\Gamma}_A(p_3)\hat{\Gamma}_A(p_1)\hat{\Gamma}_A(p_2) \Big)
  &=  \sum_{m, m'} e^{-i {\pi \over 2}(m'-m)}  e^{-2\pi i\,m\, g}  d^{F}_{m,m'}(2\pi g) \, d^{F}_{m',m}(2\pi\, g)~~,
\end{align}
so that
\begin{align}
  I &= \mathrm{tr}\Big( \hat{\Gamma}_A(p_3)[\hat{\Gamma}_A(p_2), \hat{\Gamma}_A(p_1)]\Big) \nonumber\\
    &= \sum_{m, m'}\left[e^{i {\pi \over 2}(m'-m)}  - e^{-i {\pi \over 2}(m'-m)}  \right]e^{-2\pi i\,m\, g}  d^{F}_{m,m'}(2\pi\, g) \, d^{F}_{m',m}(2\pi\, g)  \nonumber\\
    &= 2i  \sum_{m, m' = -F}^{F}\sin\left( {\pi \over 2}(m'-m)  \right) e^{-2\pi i\,m\, g}  d^{F}_{m,m'}(2\pi\, g) \, d^{F}_{m',m}(2\pi\, g) \nonumber\\
    &= \sum_{m=-F}^{F} I_m ~~.  
\end{align}
Now, we can observe that
\begin{align}
  I_{-m}
  &= 2i \sum_{m'} \sin\left( {\pi \over 2}(m'+m)  \right) e^{2\pi im\, g }  d^{F}_{-m,m'}(2\pi \, g) \,
    d^{F}_{m',-m}(2\pi \, g) \nonumber \\
  &= 2i \sum_{m'} \sin\left( {\pi \over 2}(-m'+m)  \right) e^{2\pi im\, g}  d^{F}_{-m,-m'}(2\pi \, g) \,
    d^{F}_{-m',-m}(2\pi \, g) \nonumber \\
  &= -2i \sum_{m'} \sin\left( {\pi \over 2}(m'-m)  \right) e^{2\pi im\, g}
    d^{F}_{m,m'}(2\pi \, g) \, d^{F}_{m',m}(2\pi \, g) \nonumber\\
  & = I_{m}^{*},  
\end{align}
and that (if it exists) $I_{0}$ is purely imaginary, and so must vanish. Hence, we have:
\begin{align}
  I &= \sum_{m > 0}^F (I_m + I_{-m}) \nonumber\\
    &= \sum_{m > 0}^F (I_m + I_{m}^{*}) \nonumber\\
    &= 2i\sum_{m > 0}^{F}\sum_{m' = -F}^{F}\sin\left( {\pi \over 2}(m'-m)  \right)
      \left( e^{-2\pi i\,m\,g} - e^{2\pi i\,m\,g} \right)  d^{F}_{m,m'}(2\pi \, g) \, d^{F}_{m',m}(2\pi \, g) \nonumber\\
    &= 4 \sum_{m > 0}^{F}\sum_{m' = -F}^{F} \sin\left( {\pi \over 2}(m'-m)  \right)   \sin(2\pi \, m\, g) \, d^{F}_{m,m'}(G) \, d^{F}_{m',m}(2\pi \, g) \nonumber\\
    &= 4 \sum_{m > 0}^{F}\sum_{m' = -F}^{F} (-1)^{m-m'} \sin\left( {\pi \over 2}(m'-m)  \right)
      \sin(2\pi \, m \, g) \left( d^{F}_{m,m'}(2\pi \, g) \right)^2, \label{eq:I_final} 
\end{align}
a manifestly {\sl real} quantity which can be experimentally measured. In the absence of dynamical effects, this quantity being non-zero would demonstrate a non-Abelian geometric phase, but more generally signals path-dependent evolution.

\end{widetext}

\section{Generalized Detuning}
\label{s:supp-det}

As described in the main text, uncontrolled environmental factors may contribute to unknown shifts in the experimental parameters, such as the detuning. In our case, for example, unknown ambient magnetic fields that change between calibrations may result in shifts to this detuning, $\Delta_z = \omega_{\rm RF} - \omega_{\rm Z}$.

Given a general, static detuning term in the lab frame Hamiltonian,
\begin{align}\label{eq:detuning_lab}
  \hat{H}_\Delta = \bm{\Delta} \cdot \bm{\hat{F}},
\end{align}
we can transform $\hat{H}_\Delta$ to the Floquet basis with the micromotion operator:
\begin{align}
  \label{eq:micomotion2}
  \hat{U} = \exp\left[i \Omega_0 \bm{q}(t)\cdot \hat{F}\, {\sin(\omega t) \over \omega} \right],
\end{align}
using the Baker-Hausdorff lemma:
\begin{align*}
  e^{\hat{X}} \hat{Y} e^{-\hat{X}} = {1 \over 0!} \hat{Y} + {1 \over 1!}[\hat{X}, \hat{Y}] + {1 \over 2!} \left[\hat{X} , [\hat{X}, \hat{Y}] \right] + ...~~.
\end{align*}
\begin{widetext}
    
The commutators between $\hat{H}_\Delta$ and the terms in the power series representation of $\hat{U}$ form a recurrence relation:
\begin{align*}
  [\bm{q}(t)\cdot \bm{\hat{F}}, \bm{\Delta} \cdot \bm{\hat{F}}]
  &= i ( \bm{q}(t) \times \bm{\Delta}) \cdot \bm{\hat{F}}, \\
  \left[  \bm{q}(t)\cdot \hat{\bm{F}}, [\bm{q}(t)\cdot \bm{\hat{F}}, \bm{\Delta} \cdot \bm{\hat{F}}]  \right]
  &= \bm{\Delta} \cdot \bm{\hat{F}}  - (\bm{\Delta} \cdot {\bm{q}}) (\bm{q} \cdot \bm{\hat{F}}), \\
  \left[\bm{q}(t)\cdot \bm{\hat{F}},  \left[  \bm{q}(t)\cdot \bm{\hat{F}}, [\bm{q}(t)\cdot \bm{\hat{F}}, \bm{\Delta} \cdot \bm{\hat{F}}]  \right]  \right]
  &= [\bm{q}(t)\cdot \bm{\hat{F}}, \bm{\Delta} \cdot \bm{\hat{F}}],
\end{align*}
where we have asssumed that $\bm{q \cdot q} = 1$, so that we may write
\begin{align}
  \hat{U}^\dagger \hat{H}_\Delta \hat{U}
  &= {\bm{\Delta} \cdot \bm{F}} + i(\bm{q}\times \bm{\Delta})\cdot \bm{\hat{F}} \left( i \Omega_0 {\sin \omega t \over \omega} \right)
    + {1 \over 2} \left[\bm{\Delta} \cdot \bm{\hat{F}} - (\bm{\Delta} \cdot \bm{q}) (\bm{q} \cdot \bm{\hat{F}}) \right]
    \left( i \Omega_0 {\sin \omega t \over \omega} \right)^2 \nonumber\\
  & \qquad + {i \over 3!}(\bm{q}\times \bm{\Delta})\cdot \bm{\hat{F}} \left( i \Omega_0 {\sin \omega t \over \omega} \right)^3
    + {1 \over 4!} \left[\bm{\Delta} \cdot \bm{\hat{F}} - (\bm{\Delta} \cdot \bm{q}) (\bm{q} \cdot \bm{\hat{F}}) \right]
    \left( i \Omega_0 {\sin \omega t \over \omega} \right)^4 \nonumber\\
  & \qquad + {i \over 5!}(\bm{q}\times \bm{\Delta})\cdot \bm{\hat{F}} \left( i \Omega_0 {\sin \omega t \over \omega} \right)^5
    + {1 \over 6!} \left[\bm{\Delta} \cdot \bm{\hat{F}} - (\bm{\Delta} \cdot \bm{q}) (\bm{q} \cdot \bm{\hat{F}}) \right]
    \left( i \Omega_0 {\sin \omega t \over \omega} \right)^6  + ...  \nonumber\\
  &=  {\bm{\Delta} \cdot \bm{\hat{F}}}
    + (\bm{q}\times \bm{\Delta})\cdot \bm{\hat{F}} \, \sum_{n=1}^\infty {(-1)^n \over (2n-1)!} \left( i \Omega_0 {\sin \omega t \over \omega} \right)^{2n-1}\nonumber\\
  & \qquad + \left[\bm{\Delta} \cdot \bm{\hat{F}} - (\bm{\Delta} \cdot \bm{q}) (\bm{q} \cdot \bm{\hat{F}}) \right] \sum_{n=1}^{\infty} {(-1)^n \over (2n)!}
    \left( i \Omega_0 {\sin \omega t \over \omega} \right)^{2n} \nonumber\\
  & = {\bm{\Delta} \cdot \bm{\hat{F}}}
    + (\bm{q}\times \bm{\Delta})\cdot \bm{\hat{F}} \sin\left(\Omega_0 {\sin \omega t \over \omega}  \right)
    + \left[\bm{\Delta} \cdot \bm{\hat{F}} - (\bm{\Delta} \cdot \bm{q}) (\bm{q} \cdot \bm{\hat{F}}) \right] \left[ \cos\left(\Omega_0 {\sin \omega t \over \omega}  \right) - 1 \right].
\end{align}
\end{widetext}

As in Refs. \cite{Novičenko.2019, Novičenko.2017}, we restrict our attention to the zeroth Floquet band:
\begin{align}
  \hat{H}_{\mathrm{Floq.}}^\Delta(t)
  &= {\omega \over 2\pi} \int_0^{2\pi/\omega} dt' \, (\hat{U}^\dagger \hat{H}_\Delta \hat{U})(t+t'),
\end{align}
where we take $\bm{q}$ and $\bm{\Delta}$ to be approximately static over the course of the integral. Noting that
\begin{align}
  {1 \over 2\pi} \int_0^{2\pi} d\theta \cos\left(a \sin \theta  \right) &= J_{0}(|a|), \\
   {1 \over 2\pi} \int_0^{2\pi} d\theta \sin\left(a \sin \theta  \right) &= 0~~,
\end{align}
we find that in our approximation scheme, the Floquet-transformed detuning term becomes:
\begin{align}
    \hat{H}_{\mathrm{Floq}}^\Delta(t) &= (1 - g) \bm{\Delta \cdot \hat{F}} + g \left(\bm{q \cdot \Delta}\right)(\bm{q \cdot \hat{F}}),
\end{align}
where $g =1 - J_0(\Omega_0 / \omega)$.

\section{Experimental Methods}
\label{s:supp-methods}

Our experiment uses an ensemble of ultracold neutral \Rb\ atoms in a Bose-Einstein condensate (BEC), where we prepare and manipulate a new BEC for each measurement, treating each BEC as a single quantum object (with all atoms acting in unison). To create these ultracold ensembles, we use standard laser cooling in a magneto-optical trap (MOT), and then perform forced RF-evaporative cooling in a magnetic trap, leaving atoms in the $\ket{F=2, m_F=+2}$ ground state. Atoms are then loaded into a crossed optical dipole trap (ODT) and evaporated further, until we obtain a nearly pure BEC of about $10^5$ atoms. Atoms are held in the ODT for the remaining duration of the experiment, and only released in time of flight (TOF) prior to projective measurement. 

State preparation used magnetic-dipole transitions, with microwave-frequency fields to couple  the $F=1$ and $F=2$ hyperfine manifolds, and/or RF fields to couple $m_F$ levels within a manifold. A  static magnetic bias field is applied to control the Zeeman splitting $\omega_{\rm Z}$ between $m_F$ levels, and was adjusted  so that the RF driving field  was resonant when $\omega_{\rm RF}/2\pi = 1.25$~MHz.  The microwave field is produced by mixing the output of a microwave function generator, which is detuned from the hyperfine clock transition by about 100~MHz, and an RF signal from an arbitrary waveform generator (AWG); these signals are amplified together, and transmitted through a horn antenna towards the BEC. Each of the microwave source's polarization components address different $\ket{F, \mf} \rightarrow \ket{F^{\prime}, \mf^{\prime}}$ transitions, which are frequency dependent in the presence of Zeeman splitting. By adjusting the microwave carrier frequency,   the mixed RF signal can be tuned to independently address the desired transitions~\cite{Lindon.2023}.

Through the combination of these fields, we prepare atoms in any of the three $m_F$ levels in the $F=1$ manifold with near purity (See \S\ref{s:experiment}). Preparing atoms in the $\ket{F=1, \mf=\pm 1}$ states can be achieved by a single microwave pulse, followed by a resonant RF $\pi$-pulse, in the case of the $\mf=-1$ state. For the $\ket{F=1, \mf=0}$ state, we use three separate microwave pulses transferring atoms from $\ket{F=2, \mf=+2} \rightarrow \ket{F=1, \mf=+1}$, then $\ket{F=1, \mf=+1} \rightarrow \ket{F=2, \mf=0}$, and finally $\ket{F=2, \mf=0} \rightarrow \ket{F=1, \mf=0}$.

In addition to the RF channel that mixes with the microwave carrier pulses, another AWG channel is used to generate the RF pulses for state preparation, generating holonomies, and state readout. The output from this AWG channel is amplified and sent through a pair of coils located near the BEC vacuum chamber; the oscillating current in the coils produces an RF oscillating magnetic field (Eq.~\ref{eq:mag-field}), which couples internal $m_F$ states. To achieve stronger fields after amplification,  the signal is sent through a home-built circuit to  match the impedance of the source to the transmission line and coils.

Following state preparation and the application of a holonomy (Eq.\ \ref{eq:holonomy}), we perform state readout. The ODT beams are turned off, allowing atoms to fall in TOF. During this time, a small magnetic field gradient is applied that, through the Stern-Gerlach effect, spatially separates atoms according to their $m_F$ levels. This is a projective measurement in the $\hat{F}_z$ basis, where  we obtain ensemble statistics by looking at the relative populations in each spin component through absorption imaging. To measure in other bases, we precede the TOF measurement with a short RF ``readout'' pulse with varied pulse area and phase. From a set of measurements with varied RF readout  pulses we can tomographically reconstruct the prepared state.

Prior to each set of measurements, and intermittently throughout data collection, we calibrate the RF-resonance against the background detuning by applying an RF $\pi$-pulse to a pure initial state. By measuring the final state, we can detect  large scale detunings, which  would result in imperfect population transfer. The bias static magnetic field was adjusted  to match the $\omega_{\rm RF}/2\pi = 1.25$~MHz RF carrier frequency of the driving field. This technique is limited by our ability to image the small number of atoms remaining in the initial $\mf$ state, and the short duration of a single $\pi$-pulse. To further zero the detuning we would then use the holonomy $\holonomy{\ell_1}$ in a similar way, looking for the expected state populations. This gate is more sensitive to detuning than an RF $\pi$-pulse of the same duration due to its multi-spectral decomposition. In the absence of other more sophisticated magnetometry techniques, this provided an excellent resonance calibration.

\section{Quadratic Zeeman Effect}
\label{s:supp-quadzee}

The quadratic Zeeman shift is a correction to the atomic spin energies, second order in magnetic field. In the bare spin basis (lab frame) we may write it as~\cite{Jiménez-García.2012},
\begin{equation}
    \hat{H}_{\epsilon} = \epsilon \left( \identity - \hat{F}_z^2 \right).
\end{equation}
In the case of $F=1/2$, $\hat{F}_z^2 = \identity$, hence this term can be neglected. For all other spins $F > 1/2$, $\hat{H}_{\epsilon}$ breaks the $SU(2)$ symmetry of the transformations.

The magnitude of the shift $\epsilon$ varies with the square of the applied magnetic field, $B_z$~\cite{Stamper-Kurn.2013}. In terms of the linear Zeeman splitting, $\omega_{\mathrm{Z}} = \omega_B B_z$ with $\omega_B / 2\pi \approx 0.7\ \mathrm{MHz/G}$ (in \Rb), the quadratic shift is,
\begin{equation}
    \vert \epsilon \vert = \frac{
        \left( g_s \mu_{\mathrm{B}} - g_I \mu_{\mathrm{N}} \right)^2
    }{
        E_{\mathrm{HF}} \omega_B^2 \left( 1 - 2I \right)^2
    }
    \omega_{\mathrm{Z}}^2,
\end{equation}
where $g_s$ is the $g$-factor of the electron, and $g_I \mu_{\mathrm{N}} / \hbar$ is the nuclear gyromagnetic ratio. For \Rb, $E_{\mathrm{HF}} / h \approx 6.835\ \mathrm{GHz}$ is the ground state hyperfine splitting, $I = 3/2$ is the nuclear spin. For the Zeeman splitting used here, $\omega_{\mathrm{Z}} / 2\pi = 1.25\ \mathrm{MHz}$, we find $\epsilon / 2\pi \approx 0.228\ \mathrm{kHz}$. The detunings (Eq.~\ref{eq:delta-ham}) observed here are of the order $\Delta_z / 2\pi \leq 0.8\ \mathrm{kHz}$, so we assume that $\epsilon$ is constant across all measurements.

\section{Numerics}
\label{s:supp-numerics}

All data processing and theory calculations were done in \texttt{Julia} ~\cite{Julia-2017}. For comparisons of data to theory, we rewrite the Schr\"odinger equation in terms of the evolution operator $\hat{U}$,
\begin{equation}
    \partial_t \hat{U}(t) = -i \hat{H}(t) \hat{U}(t),
    \label{eq:schro-evolution}
\end{equation}
where $\hat{H}$ is the Hamiltonian of interest, and the initial condition is $\hat{U}(t=0) = \identity$; this is numerically integrated using the \texttt{OrdinaryDiffEq.jl} package~\cite{DifferentialEquations.jl-2017}. All parameters used in simulation were fixed to the values measured in experiment, with only the detuning $\Delta_z$ (Eq.~\ref{eq:delta-ham}) permitted to vary; these fits were performed using the Nelder-Mead non-linear optimization algorithm with the norm-squared difference as the objective function, using \texttt{Optim.jl}~\cite{Optim.jl-2018}.

In order to fit the holonomies (Eq.~\ref{eq:holonomy}) to a set of tomographic measurements (see Fig.~\ref{fig:op-fid}), we decomposed the holonomy into a general operator in $SU(3)$, parameterized through the Gell-Mann matrices,
\begin{equation}
    \hat{\Gamma}_{A} (\ell) = \exp{\left( -i \sum_{n=1}^8 c_n \hat{\lambda}_n \right)}.
\end{equation}
For each set of tomographic measurements, the coefficients $c_n$ of each generator $\hat{\lambda}_n$ were fit to the data; therefore, these fits have no information about the Hamiltonian which produced the transformation, only the resulting set of projections. These measured holonomies could then be compared to theoretical predictions, computed by the numerical integration of Eq.~\ref{eq:schro-evolution} above, in order to produce the fidelities shown in Fig.~\ref{fig:op-fid}.

% References
%apsrev4-2.bst 2019-01-14 (MD) hand-edited version of apsrev4-1.bst
%Control: key (0)
%Control: author (8) initials jnrlst
%Control: editor formatted (1) identically to author
%Control: production of article title (0) allowed
%Control: page (0) single
%Control: year (1) truncated
%Control: production of eprint (1) enabled
%


\begin{thebibliography}{63}%
\makeatletter
\providecommand \@ifxundefined [1]{%
 \@ifx{#1\undefined}
}%
\providecommand \@ifnum [1]{%
 \ifnum #1\expandafter \@firstoftwo
 \else \expandafter \@secondoftwo
 \fi
}%
\providecommand \@ifx [1]{%
 \ifx #1\expandafter \@firstoftwo
 \else \expandafter \@secondoftwo
 \fi
}%
\providecommand \natexlab [1]{#1}%
\providecommand \enquote  [1]{``#1''}%
\providecommand \bibnamefont  [1]{#1}%
\providecommand \bibfnamefont [1]{#1}%
\providecommand \citenamefont [1]{#1}%
\providecommand \href@noop [0]{\@secondoftwo}%
\providecommand \href [0]{\begingroup \@sanitize@url \@href}%
\providecommand \@href[1]{\@@startlink{#1}\@@href}%
\providecommand \@@href[1]{\endgroup#1\@@endlink}%
\providecommand \@sanitize@url [0]{\catcode `\\12\catcode `\$12\catcode
  `\&12\catcode `\#12\catcode `\^12\catcode `\_12\catcode `\%12\relax}%
\providecommand \@@startlink[1]{}%
\providecommand \@@endlink[0]{}%
\providecommand \url  [0]{\begingroup\@sanitize@url \@url }%
\providecommand \@url [1]{\endgroup\@href {#1}{\urlprefix }}%
\providecommand \urlprefix  [0]{URL }%
\providecommand \Eprint [0]{\href }%
\providecommand \doibase [0]{https://doi.org/}%
\providecommand \selectlanguage [0]{\@gobble}%
\providecommand \bibinfo  [0]{\@secondoftwo}%
\providecommand \bibfield  [0]{\@secondoftwo}%
\providecommand \translation [1]{[#1]}%
\providecommand \BibitemOpen [0]{}%
\providecommand \bibitemStop [0]{}%
\providecommand \bibitemNoStop [0]{.\EOS\space}%
\providecommand \EOS [0]{\spacefactor3000\relax}%
\providecommand \BibitemShut  [1]{\csname bibitem#1\endcsname}%
\let\auto@bib@innerbib\@empty
%</preamble>
\bibitem [{\citenamefont {DiVincenzo}(2000)}]{DiVincenzo.2000}%
  \BibitemOpen
  \bibfield  {author} {\bibinfo {author} {\bibfnamefont {D.~P.}\ \bibnamefont
  {DiVincenzo}},\ }\bibfield  {title} {\bibinfo {title} {{The Physical
  Implementation of Quantum Computation}},\ }\href
  {https://doi.org/10.1002/1521-3978(200009)48:9/11<771::aid-prop771>3.0.co;2-e}
  {\bibfield  {journal} {\bibinfo  {journal} {Fortschritte der Physik}\
  }\textbf {\bibinfo {volume} {48}},\ \bibinfo {pages} {771} (\bibinfo {year}
  {2000})}\BibitemShut {NoStop}%
\bibitem [{\citenamefont {Devitt}\ \emph {et~al.}(2013)\citenamefont {Devitt},
  \citenamefont {Munro},\ and\ \citenamefont {Nemoto}}]{Devitt.2013}%
  \BibitemOpen
  \bibfield  {author} {\bibinfo {author} {\bibfnamefont {S.~J.}\ \bibnamefont
  {Devitt}}, \bibinfo {author} {\bibfnamefont {W.~J.}\ \bibnamefont {Munro}},\
  and\ \bibinfo {author} {\bibfnamefont {K.}~\bibnamefont {Nemoto}},\
  }\bibfield  {title} {\bibinfo {title} {{Quantum error correction for
  beginners}},\ }\href {https://doi.org/10.1088/0034-4885/76/7/076001}
  {\bibfield  {journal} {\bibinfo  {journal} {Reports on Progress in Physics}\
  }\textbf {\bibinfo {volume} {76}},\ \bibinfo {pages} {076001} (\bibinfo
  {year} {2013})}\BibitemShut {NoStop}%
\bibitem [{\citenamefont {Pachos}\ and\ \citenamefont
  {Zanardi}(2001)}]{Pachos.2001}%
  \BibitemOpen
  \bibfield  {author} {\bibinfo {author} {\bibfnamefont {J.}~\bibnamefont
  {Pachos}}\ and\ \bibinfo {author} {\bibfnamefont {P.}~\bibnamefont
  {Zanardi}},\ }\bibfield  {title} {\bibinfo {title} {{Quantum Holonomies for
  Quantum Computing}},\ }\href {https://doi.org/10.1142/s0217979201004836}
  {\bibfield  {journal} {\bibinfo  {journal} {International Journal of Modern
  Physics B}\ }\textbf {\bibinfo {volume} {15}},\ \bibinfo {pages} {1257}
  (\bibinfo {year} {2001})}\BibitemShut {NoStop}%
\bibitem [{\citenamefont {Zanardi}\ and\ \citenamefont
  {Rasetti}(1999)}]{Zanardi.1999qb}%
  \BibitemOpen
  \bibfield  {author} {\bibinfo {author} {\bibfnamefont {P.}~\bibnamefont
  {Zanardi}}\ and\ \bibinfo {author} {\bibfnamefont {M.}~\bibnamefont
  {Rasetti}},\ }\bibfield  {title} {\bibinfo {title} {{Holonomic quantum
  computation}},\ }\href {https://doi.org/10.1016/s0375-9601(99)00803-8}
  {\bibfield  {journal} {\bibinfo  {journal} {Physics Letters A}\ }\textbf
  {\bibinfo {volume} {264}},\ \bibinfo {pages} {94} (\bibinfo {year}
  {1999})}\BibitemShut {NoStop}%
\bibitem [{\citenamefont {Pachos}\ \emph {et~al.}(1999)\citenamefont {Pachos},
  \citenamefont {Zanardi},\ and\ \citenamefont {Rasetti}}]{Pachos.2000}%
  \BibitemOpen
  \bibfield  {author} {\bibinfo {author} {\bibfnamefont {J.}~\bibnamefont
  {Pachos}}, \bibinfo {author} {\bibfnamefont {P.}~\bibnamefont {Zanardi}},\
  and\ \bibinfo {author} {\bibfnamefont {M.}~\bibnamefont {Rasetti}},\
  }\bibfield  {title} {\bibinfo {title} {{Non-Abelian Berry connections for
  quantum computation}},\ }\href {https://doi.org/10.1103/physreva.61.010305}
  {\bibfield  {journal} {\bibinfo  {journal} {Physical Review A}\ }\textbf
  {\bibinfo {volume} {61}},\ \bibinfo {pages} {010305(R)} (\bibinfo {year}
  {1999})}\BibitemShut {NoStop}%
\bibitem [{\citenamefont {Colmenar}\ \emph {et~al.}(2022)\citenamefont
  {Colmenar}, \citenamefont {G\"{u}ng\"{o}rd\"{u}},\ and\ \citenamefont
  {Kestner}}]{Colmenar.2022}%
  \BibitemOpen
  \bibfield  {author} {\bibinfo {author} {\bibfnamefont {R.~K.~L.}\
  \bibnamefont {Colmenar}}, \bibinfo {author} {\bibfnamefont {U.}~\bibnamefont
  {G\"{u}ng\"{o}rd\"{u}}},\ and\ \bibinfo {author} {\bibfnamefont {J.~P.}\
  \bibnamefont {Kestner}},\ }\bibfield  {title} {\bibinfo {title} {{Conditions
  for Equivalent Noise Sensitivity of Geometric and Dynamical Quantum Gates}},\
  }\href {https://doi.org/10.1103/prxquantum.3.030310} {\bibfield  {journal}
  {\bibinfo  {journal} {PRX Quantum}\ }\textbf {\bibinfo {volume} {3}},\
  \bibinfo {pages} {030310} (\bibinfo {year} {2022})}\BibitemShut {NoStop}%
\bibitem [{Note1()}]{Note1}%
  \BibitemOpen
  \bibinfo {note} {Also referred to as an \emph {anholonomy} in earlier work on
  geometric phase.}\BibitemShut {Stop}%
\bibitem [{\citenamefont {Berry}(1984)}]{berry1984quantal}%
  \BibitemOpen
  \bibfield  {author} {\bibinfo {author} {\bibfnamefont {M.~V.}\ \bibnamefont
  {Berry}},\ }\bibfield  {title} {\bibinfo {title} {Quantal phase factors
  accompanying adiabatic changes},\ }\href
  {https://doi.org/10.1098/rspa.1984.0023} {\bibfield  {journal} {\bibinfo
  {journal} {Proceedings of the Royal Society of London. A. Mathematical and
  Physical Sciences}\ }\textbf {\bibinfo {volume} {392}},\ \bibinfo {pages}
  {45} (\bibinfo {year} {1984})}\BibitemShut {NoStop}%
\bibitem [{\citenamefont {Zhang}\ \emph {et~al.}(2021)\citenamefont {Zhang},
  \citenamefont {Kyaw}, \citenamefont {Filipp}, \citenamefont {Kwek},
  \citenamefont {Sj\"{o}qvist},\ and\ \citenamefont {Tong}}]{Zhang.20212cna}%
  \BibitemOpen
  \bibfield  {author} {\bibinfo {author} {\bibfnamefont {J.}~\bibnamefont
  {Zhang}}, \bibinfo {author} {\bibfnamefont {T.~H.}\ \bibnamefont {Kyaw}},
  \bibinfo {author} {\bibfnamefont {S.}~\bibnamefont {Filipp}}, \bibinfo
  {author} {\bibfnamefont {L.-C.}\ \bibnamefont {Kwek}}, \bibinfo {author}
  {\bibfnamefont {E.}~\bibnamefont {Sj\"{o}qvist}},\ and\ \bibinfo {author}
  {\bibfnamefont {D.}~\bibnamefont {Tong}},\ }\bibfield  {title} {\bibinfo
  {title} {{Geometric and holonomic quantum computation}},\ }\bibfield
  {journal} {\bibinfo  {journal} {arXiv}\ }\href
  {https://doi.org/10.48550/arxiv.2110.03602} {10.48550/arxiv.2110.03602}
  (\bibinfo {year} {2021})\BibitemShut {NoStop}%
\bibitem [{\citenamefont {Sj\"{o}qvist}\ \emph {et~al.}(2012)\citenamefont
  {Sj\"{o}qvist}, \citenamefont {Tong}, \citenamefont {Andersson},
  \citenamefont {Hessmo}, \citenamefont {Johansson},\ and\ \citenamefont
  {Singh}}]{Sjöqvist.2012}%
  \BibitemOpen
  \bibfield  {author} {\bibinfo {author} {\bibfnamefont {E.}~\bibnamefont
  {Sj\"{o}qvist}}, \bibinfo {author} {\bibfnamefont {D.~M.}\ \bibnamefont
  {Tong}}, \bibinfo {author} {\bibfnamefont {L.~M.}\ \bibnamefont {Andersson}},
  \bibinfo {author} {\bibfnamefont {B.}~\bibnamefont {Hessmo}}, \bibinfo
  {author} {\bibfnamefont {M.}~\bibnamefont {Johansson}},\ and\ \bibinfo
  {author} {\bibfnamefont {K.}~\bibnamefont {Singh}},\ }\bibfield  {title}
  {\bibinfo {title} {{Non-adiabatic holonomic quantum computation}},\ }\href
  {https://doi.org/10.1088/1367-2630/14/10/103035} {\bibfield  {journal}
  {\bibinfo  {journal} {New Journal of Physics}\ }\textbf {\bibinfo {volume}
  {14}},\ \bibinfo {pages} {103035} (\bibinfo {year} {2012})}\BibitemShut
  {NoStop}%
\bibitem [{\citenamefont {Duan}\ \emph {et~al.}(2001)\citenamefont {Duan},
  \citenamefont {Cirac},\ and\ \citenamefont {Zoller}}]{Duan.2001}%
  \BibitemOpen
  \bibfield  {author} {\bibinfo {author} {\bibfnamefont {L.-M.}\ \bibnamefont
  {Duan}}, \bibinfo {author} {\bibfnamefont {J.~I.}\ \bibnamefont {Cirac}},\
  and\ \bibinfo {author} {\bibfnamefont {P.}~\bibnamefont {Zoller}},\
  }\bibfield  {title} {\bibinfo {title} {{Geometric Manipulation of Trapped
  Ions for Quantum Computation}},\ }\href
  {https://doi.org/10.1126/science.1058835} {\bibfield  {journal} {\bibinfo
  {journal} {Science}\ }\textbf {\bibinfo {volume} {292}},\ \bibinfo {pages}
  {1695} (\bibinfo {year} {2001})}\BibitemShut {NoStop}%
\bibitem [{\citenamefont {Ai}\ \emph {et~al.}(2022)\citenamefont {Ai},
  \citenamefont {Li}, \citenamefont {He}, \citenamefont {Xue}, \citenamefont
  {Cui}, \citenamefont {Huang}, \citenamefont {Li},\ and\ \citenamefont
  {Guo}}]{Ai.2022}%
  \BibitemOpen
  \bibfield  {author} {\bibinfo {author} {\bibfnamefont {M.-Z.}\ \bibnamefont
  {Ai}}, \bibinfo {author} {\bibfnamefont {S.}~\bibnamefont {Li}}, \bibinfo
  {author} {\bibfnamefont {R.}~\bibnamefont {He}}, \bibinfo {author}
  {\bibfnamefont {Z.-Y.}\ \bibnamefont {Xue}}, \bibinfo {author} {\bibfnamefont
  {J.-M.}\ \bibnamefont {Cui}}, \bibinfo {author} {\bibfnamefont {Y.-F.}\
  \bibnamefont {Huang}}, \bibinfo {author} {\bibfnamefont {C.-F.}\ \bibnamefont
  {Li}},\ and\ \bibinfo {author} {\bibfnamefont {G.-C.}\ \bibnamefont {Guo}},\
  }\bibfield  {title} {\bibinfo {title} {{Experimental realization of
  nonadiabatic holonomic single‐qubit quantum gates with two dark paths in a
  trapped ion}},\ }\href {https://doi.org/10.1016/j.fmre.2021.11.031}
  {\bibfield  {journal} {\bibinfo  {journal} {Fundamental Research}\ }\textbf
  {\bibinfo {volume} {2}},\ \bibinfo {pages} {661} (\bibinfo {year}
  {2022})}\BibitemShut {NoStop}%
\bibitem [{\citenamefont {Shui}\ \emph {et~al.}(2021)\citenamefont {Shui},
  \citenamefont {Jin}, \citenamefont {Li}, \citenamefont {Wei}, \citenamefont
  {Chen}, \citenamefont {Li},\ and\ \citenamefont {Zhou}}]{Shui.2021}%
  \BibitemOpen
  \bibfield  {author} {\bibinfo {author} {\bibfnamefont {H.}~\bibnamefont
  {Shui}}, \bibinfo {author} {\bibfnamefont {S.}~\bibnamefont {Jin}}, \bibinfo
  {author} {\bibfnamefont {Z.}~\bibnamefont {Li}}, \bibinfo {author}
  {\bibfnamefont {F.}~\bibnamefont {Wei}}, \bibinfo {author} {\bibfnamefont
  {X.}~\bibnamefont {Chen}}, \bibinfo {author} {\bibfnamefont {X.}~\bibnamefont
  {Li}},\ and\ \bibinfo {author} {\bibfnamefont {X.}~\bibnamefont {Zhou}},\
  }\bibfield  {title} {\bibinfo {title} {{Atom-orbital qubit under nonadiabatic
  holonomic quantum control}},\ }\href
  {https://doi.org/10.1103/physreva.104.l060601} {\bibfield  {journal}
  {\bibinfo  {journal} {Physical Review A}\ }\textbf {\bibinfo {volume}
  {104}},\ \bibinfo {pages} {L060601} (\bibinfo {year} {2021})}\BibitemShut
  {NoStop}%
\bibitem [{\citenamefont {Leroux}\ \emph {et~al.}(2018)\citenamefont {Leroux},
  \citenamefont {Pandey}, \citenamefont {Rehbi}, \citenamefont {Chevy},
  \citenamefont {Miniatura}, \citenamefont {Grémaud},\ and\ \citenamefont
  {Wilkowski}}]{Leroux.2018}%
  \BibitemOpen
  \bibfield  {author} {\bibinfo {author} {\bibfnamefont {F.}~\bibnamefont
  {Leroux}}, \bibinfo {author} {\bibfnamefont {K.}~\bibnamefont {Pandey}},
  \bibinfo {author} {\bibfnamefont {R.}~\bibnamefont {Rehbi}}, \bibinfo
  {author} {\bibfnamefont {F.}~\bibnamefont {Chevy}}, \bibinfo {author}
  {\bibfnamefont {C.}~\bibnamefont {Miniatura}}, \bibinfo {author}
  {\bibfnamefont {B.}~\bibnamefont {Grémaud}},\ and\ \bibinfo {author}
  {\bibfnamefont {D.}~\bibnamefont {Wilkowski}},\ }\bibfield  {title} {\bibinfo
  {title} {{Non-Abelian adiabatic geometric transformations in a cold strontium
  gas}},\ }\href {https://doi.org/10.1038/s41467-018-05865-3} {\bibfield
  {journal} {\bibinfo  {journal} {Nature Communications}\ }\textbf {\bibinfo
  {volume} {9}},\ \bibinfo {pages} {3580} (\bibinfo {year} {2018})}\BibitemShut
  {NoStop}%
\bibitem [{\citenamefont {Feng}\ \emph {et~al.}(2013)\citenamefont {Feng},
  \citenamefont {Xu},\ and\ \citenamefont {Long}}]{Feng.2013}%
  \BibitemOpen
  \bibfield  {author} {\bibinfo {author} {\bibfnamefont {G.}~\bibnamefont
  {Feng}}, \bibinfo {author} {\bibfnamefont {G.}~\bibnamefont {Xu}},\ and\
  \bibinfo {author} {\bibfnamefont {G.}~\bibnamefont {Long}},\ }\bibfield
  {title} {\bibinfo {title} {{Experimental Realization of Nonadiabatic
  Holonomic Quantum Computation}},\ }\href
  {https://doi.org/10.1103/physrevlett.110.190501} {\bibfield  {journal}
  {\bibinfo  {journal} {Physical Review Letters}\ }\textbf {\bibinfo {volume}
  {110}},\ \bibinfo {pages} {190501} (\bibinfo {year} {2013})}\BibitemShut
  {NoStop}%
\bibitem [{\citenamefont {Jones}\ \emph {et~al.}(2000)\citenamefont {Jones},
  \citenamefont {Vedral}, \citenamefont {Ekert},\ and\ \citenamefont
  {Castagnoli}}]{Jones.2000}%
  \BibitemOpen
  \bibfield  {author} {\bibinfo {author} {\bibfnamefont {J.~A.}\ \bibnamefont
  {Jones}}, \bibinfo {author} {\bibfnamefont {V.}~\bibnamefont {Vedral}},
  \bibinfo {author} {\bibfnamefont {A.}~\bibnamefont {Ekert}},\ and\ \bibinfo
  {author} {\bibfnamefont {G.}~\bibnamefont {Castagnoli}},\ }\bibfield  {title}
  {\bibinfo {title} {{Geometric quantum computation using nuclear magnetic
  resonance}},\ }\href {https://doi.org/10.1038/35002528} {\bibfield  {journal}
  {\bibinfo  {journal} {Nature}\ }\textbf {\bibinfo {volume} {403}},\ \bibinfo
  {pages} {869} (\bibinfo {year} {2000})}\BibitemShut {NoStop}%
\bibitem [{\citenamefont {Xu}\ \emph {et~al.}(2022)\citenamefont {Xu},
  \citenamefont {Sun}, \citenamefont {Wei}, \citenamefont {Du}, \citenamefont
  {Luo}, \citenamefont {Yan}, \citenamefont {Feng},\ and\ \citenamefont
  {Su}}]{Xu.2022q69}%
  \BibitemOpen
  \bibfield  {author} {\bibinfo {author} {\bibfnamefont {J.-Z.}\ \bibnamefont
  {Xu}}, \bibinfo {author} {\bibfnamefont {L.-N.}\ \bibnamefont {Sun}},
  \bibinfo {author} {\bibfnamefont {J.-F.}\ \bibnamefont {Wei}}, \bibinfo
  {author} {\bibfnamefont {Y.-L.}\ \bibnamefont {Du}}, \bibinfo {author}
  {\bibfnamefont {R.}~\bibnamefont {Luo}}, \bibinfo {author} {\bibfnamefont
  {L.-L.}\ \bibnamefont {Yan}}, \bibinfo {author} {\bibfnamefont
  {M.}~\bibnamefont {Feng}},\ and\ \bibinfo {author} {\bibfnamefont {S.-L.}\
  \bibnamefont {Su}},\ }\bibfield  {title} {\bibinfo {title} {{Two-Qubit
  Geometric Gates Based on Ground-State Blockade of Rydberg Atoms}},\ }\href
  {https://doi.org/10.1088/0256-307x/39/9/090301} {\bibfield  {journal}
  {\bibinfo  {journal} {Chinese Physics Letters}\ }\textbf {\bibinfo {volume}
  {39}},\ \bibinfo {pages} {090301} (\bibinfo {year} {2022})}\BibitemShut
  {NoStop}%
\bibitem [{\citenamefont {Yan}\ \emph {et~al.}(2019)\citenamefont {Yan},
  \citenamefont {Liu}, \citenamefont {Xu}, \citenamefont {Song}, \citenamefont
  {Liu}, \citenamefont {Zhang}, \citenamefont {Deng}, \citenamefont {Yan},
  \citenamefont {Rong}, \citenamefont {Huang}, \citenamefont {Yung},
  \citenamefont {Chen},\ and\ \citenamefont {Yu}}]{Yan.2019}%
  \BibitemOpen
  \bibfield  {author} {\bibinfo {author} {\bibfnamefont {T.}~\bibnamefont
  {Yan}}, \bibinfo {author} {\bibfnamefont {B.-J.}\ \bibnamefont {Liu}},
  \bibinfo {author} {\bibfnamefont {K.}~\bibnamefont {Xu}}, \bibinfo {author}
  {\bibfnamefont {C.}~\bibnamefont {Song}}, \bibinfo {author} {\bibfnamefont
  {S.}~\bibnamefont {Liu}}, \bibinfo {author} {\bibfnamefont {Z.}~\bibnamefont
  {Zhang}}, \bibinfo {author} {\bibfnamefont {H.}~\bibnamefont {Deng}},
  \bibinfo {author} {\bibfnamefont {Z.}~\bibnamefont {Yan}}, \bibinfo {author}
  {\bibfnamefont {H.}~\bibnamefont {Rong}}, \bibinfo {author} {\bibfnamefont
  {K.}~\bibnamefont {Huang}}, \bibinfo {author} {\bibfnamefont {M.-H.}\
  \bibnamefont {Yung}}, \bibinfo {author} {\bibfnamefont {Y.}~\bibnamefont
  {Chen}},\ and\ \bibinfo {author} {\bibfnamefont {D.}~\bibnamefont {Yu}},\
  }\bibfield  {title} {\bibinfo {title} {{Experimental Realization of
  Nonadiabatic Shortcut to Non-Abelian Geometric Gates}},\ }\href
  {https://doi.org/10.1103/physrevlett.122.080501} {\bibfield  {journal}
  {\bibinfo  {journal} {Physical Review Letters}\ }\textbf {\bibinfo {volume}
  {122}},\ \bibinfo {pages} {080501} (\bibinfo {year} {2019})}\BibitemShut
  {NoStop}%
\bibitem [{\citenamefont {Li}\ \emph {et~al.}(2022)\citenamefont {Li},
  \citenamefont {Dong}, \citenamefont {Zheng}, \citenamefont {Zhang},
  \citenamefont {Ma}, \citenamefont {Liu}, \citenamefont {Wang}, \citenamefont
  {Li}, \citenamefont {Liu}, \citenamefont {Zhao}, \citenamefont {Lan},
  \citenamefont {Li}, \citenamefont {Tan},\ and\ \citenamefont {Yu}}]{Li.2022}%
  \BibitemOpen
  \bibfield  {author} {\bibinfo {author} {\bibfnamefont {Y.}~\bibnamefont
  {Li}}, \bibinfo {author} {\bibfnamefont {Y.}~\bibnamefont {Dong}}, \bibinfo
  {author} {\bibfnamefont {W.}~\bibnamefont {Zheng}}, \bibinfo {author}
  {\bibfnamefont {Y.}~\bibnamefont {Zhang}}, \bibinfo {author} {\bibfnamefont
  {Z.}~\bibnamefont {Ma}}, \bibinfo {author} {\bibfnamefont {Q.}~\bibnamefont
  {Liu}}, \bibinfo {author} {\bibfnamefont {J.}~\bibnamefont {Wang}}, \bibinfo
  {author} {\bibfnamefont {Y.}~\bibnamefont {Li}}, \bibinfo {author}
  {\bibfnamefont {Y.}~\bibnamefont {Liu}}, \bibinfo {author} {\bibfnamefont
  {J.}~\bibnamefont {Zhao}}, \bibinfo {author} {\bibfnamefont {D.}~\bibnamefont
  {Lan}}, \bibinfo {author} {\bibfnamefont {S.}~\bibnamefont {Li}}, \bibinfo
  {author} {\bibfnamefont {X.}~\bibnamefont {Tan}},\ and\ \bibinfo {author}
  {\bibfnamefont {Y.}~\bibnamefont {Yu}},\ }\bibfield  {title} {\bibinfo
  {title} {{Nonadiabatic Geometric Gates with a Shortened Loop in a
  Superconducting Circuit}},\ }\href {https://doi.org/10.1002/pssb.202200040}
  {\bibfield  {journal} {\bibinfo  {journal} {physica status solidi (b)}\
  }\textbf {\bibinfo {volume} {259}},\ \bibinfo {pages} {2200040} (\bibinfo
  {year} {2022})}\BibitemShut {NoStop}%
\bibitem [{\citenamefont {Yang}\ \emph {et~al.}(2023)\citenamefont {Yang},
  \citenamefont {Guo}, \citenamefont {Zhang}, \citenamefont {Du}, \citenamefont
  {Zhang}, \citenamefont {Tao}, \citenamefont {Chen}, \citenamefont {Duan},
  \citenamefont {Jia}, \citenamefont {Kong},\ and\ \citenamefont
  {Guo}}]{Yang.2023}%
  \BibitemOpen
  \bibfield  {author} {\bibinfo {author} {\bibfnamefont {X.-X.}\ \bibnamefont
  {Yang}}, \bibinfo {author} {\bibfnamefont {L.-L.}\ \bibnamefont {Guo}},
  \bibinfo {author} {\bibfnamefont {H.-F.}\ \bibnamefont {Zhang}}, \bibinfo
  {author} {\bibfnamefont {L.}~\bibnamefont {Du}}, \bibinfo {author}
  {\bibfnamefont {C.}~\bibnamefont {Zhang}}, \bibinfo {author} {\bibfnamefont
  {H.-R.}\ \bibnamefont {Tao}}, \bibinfo {author} {\bibfnamefont
  {Y.}~\bibnamefont {Chen}}, \bibinfo {author} {\bibfnamefont {P.}~\bibnamefont
  {Duan}}, \bibinfo {author} {\bibfnamefont {Z.-L.}\ \bibnamefont {Jia}},
  \bibinfo {author} {\bibfnamefont {W.-C.}\ \bibnamefont {Kong}},\ and\
  \bibinfo {author} {\bibfnamefont {G.-P.}\ \bibnamefont {Guo}},\ }\bibfield
  {title} {\bibinfo {title} {{Experimental Implementation of Short-Path
  Nonadiabatic Geometric Gates in a Superconducting Circuit}},\ }\href
  {https://doi.org/10.1103/physrevapplied.19.044076} {\bibfield  {journal}
  {\bibinfo  {journal} {Physical Review Applied}\ }\textbf {\bibinfo {volume}
  {19}},\ \bibinfo {pages} {044076} (\bibinfo {year} {2023})}\BibitemShut
  {NoStop}%
\bibitem [{\citenamefont {Abdumalikov~Jr}\ \emph {et~al.}(2013)\citenamefont
  {Abdumalikov~Jr}, \citenamefont {Fink}, \citenamefont {Juliusson},
  \citenamefont {Pechal}, \citenamefont {Berger}, \citenamefont {Wallraff},\
  and\ \citenamefont {Filipp}}]{Jr.2013}%
  \BibitemOpen
  \bibfield  {author} {\bibinfo {author} {\bibfnamefont {A.~A.}\ \bibnamefont
  {Abdumalikov~Jr}}, \bibinfo {author} {\bibfnamefont {J.~M.}\ \bibnamefont
  {Fink}}, \bibinfo {author} {\bibfnamefont {K.}~\bibnamefont {Juliusson}},
  \bibinfo {author} {\bibfnamefont {M.}~\bibnamefont {Pechal}}, \bibinfo
  {author} {\bibfnamefont {S.}~\bibnamefont {Berger}}, \bibinfo {author}
  {\bibfnamefont {A.}~\bibnamefont {Wallraff}},\ and\ \bibinfo {author}
  {\bibfnamefont {S.}~\bibnamefont {Filipp}},\ }\bibfield  {title} {\bibinfo
  {title} {{Experimental realization of non-Abelian non-adiabatic geometric
  gates}},\ }\href {https://doi.org/10.1038/nature12010} {\bibfield  {journal}
  {\bibinfo  {journal} {Nature}\ }\textbf {\bibinfo {volume} {496}},\ \bibinfo
  {pages} {482} (\bibinfo {year} {2013})}\BibitemShut {NoStop}%
\bibitem [{\citenamefont {Pinske}\ \emph {et~al.}(2020)\citenamefont {Pinske},
  \citenamefont {Teuber},\ and\ \citenamefont {Scheel}}]{Pinske.2020}%
  \BibitemOpen
  \bibfield  {author} {\bibinfo {author} {\bibfnamefont {J.}~\bibnamefont
  {Pinske}}, \bibinfo {author} {\bibfnamefont {L.}~\bibnamefont {Teuber}},\
  and\ \bibinfo {author} {\bibfnamefont {S.}~\bibnamefont {Scheel}},\
  }\bibfield  {title} {\bibinfo {title} {{Highly degenerate photonic waveguide
  structures for holonomic computation}},\ }\href
  {https://doi.org/10.1103/physreva.101.062314} {\bibfield  {journal} {\bibinfo
   {journal} {Physical Review A}\ }\textbf {\bibinfo {volume} {101}},\ \bibinfo
  {pages} {062314} (\bibinfo {year} {2020})}\BibitemShut {NoStop}%
\bibitem [{\citenamefont {Kremer}\ \emph {et~al.}(2019)\citenamefont {Kremer},
  \citenamefont {Teuber}, \citenamefont {Szameit},\ and\ \citenamefont
  {Scheel}}]{Kremer.2019}%
  \BibitemOpen
  \bibfield  {author} {\bibinfo {author} {\bibfnamefont {M.}~\bibnamefont
  {Kremer}}, \bibinfo {author} {\bibfnamefont {L.}~\bibnamefont {Teuber}},
  \bibinfo {author} {\bibfnamefont {A.}~\bibnamefont {Szameit}},\ and\ \bibinfo
  {author} {\bibfnamefont {S.}~\bibnamefont {Scheel}},\ }\bibfield  {title}
  {\bibinfo {title} {{Optimal design strategy for non-Abelian geometric phases
  using Abelian gauge fields based on quantum metric}},\ }\href
  {https://doi.org/10.1103/physrevresearch.1.033117} {\bibfield  {journal}
  {\bibinfo  {journal} {Physical Review Research}\ }\textbf {\bibinfo {volume}
  {1}},\ \bibinfo {pages} {033117} (\bibinfo {year} {2019})}\BibitemShut
  {NoStop}%
\bibitem [{\citenamefont {Sekiguchi}\ \emph {et~al.}(2017)\citenamefont
  {Sekiguchi}, \citenamefont {Niikura}, \citenamefont {Kuroiwa}, \citenamefont
  {Kano},\ and\ \citenamefont {Kosaka}}]{Sekiguchi.2017}%
  \BibitemOpen
  \bibfield  {author} {\bibinfo {author} {\bibfnamefont {Y.}~\bibnamefont
  {Sekiguchi}}, \bibinfo {author} {\bibfnamefont {N.}~\bibnamefont {Niikura}},
  \bibinfo {author} {\bibfnamefont {R.}~\bibnamefont {Kuroiwa}}, \bibinfo
  {author} {\bibfnamefont {H.}~\bibnamefont {Kano}},\ and\ \bibinfo {author}
  {\bibfnamefont {H.}~\bibnamefont {Kosaka}},\ }\bibfield  {title} {\bibinfo
  {title} {{Optical holonomic single quantum gates with a geometric spin under
  a zero field}},\ }\href {https://doi.org/10.1038/nphoton.2017.40} {\bibfield
  {journal} {\bibinfo  {journal} {Nature Photonics}\ }\textbf {\bibinfo
  {volume} {11}},\ \bibinfo {pages} {309} (\bibinfo {year} {2017})}\BibitemShut
  {NoStop}%
\bibitem [{\citenamefont {Nagata}\ \emph {et~al.}(2018)\citenamefont {Nagata},
  \citenamefont {Kuramitani}, \citenamefont {Sekiguchi},\ and\ \citenamefont
  {Kosaka}}]{Nagata.2018}%
  \BibitemOpen
  \bibfield  {author} {\bibinfo {author} {\bibfnamefont {K.}~\bibnamefont
  {Nagata}}, \bibinfo {author} {\bibfnamefont {K.}~\bibnamefont {Kuramitani}},
  \bibinfo {author} {\bibfnamefont {Y.}~\bibnamefont {Sekiguchi}},\ and\
  \bibinfo {author} {\bibfnamefont {H.}~\bibnamefont {Kosaka}},\ }\bibfield
  {title} {\bibinfo {title} {{Universal holonomic quantum gates over geometric
  spin qubits with polarised microwaves}},\ }\href
  {https://doi.org/10.1038/s41467-018-05664-w} {\bibfield  {journal} {\bibinfo
  {journal} {Nature Communications}\ }\textbf {\bibinfo {volume} {9}},\
  \bibinfo {pages} {3227} (\bibinfo {year} {2018})}\BibitemShut {NoStop}%
\bibitem [{\citenamefont {Zu}\ \emph {et~al.}(2014)\citenamefont {Zu},
  \citenamefont {Wang}, \citenamefont {He}, \citenamefont {Zhang},
  \citenamefont {Dai}, \citenamefont {Wang},\ and\ \citenamefont
  {Duan}}]{Zu.2014}%
  \BibitemOpen
  \bibfield  {author} {\bibinfo {author} {\bibfnamefont {C.}~\bibnamefont
  {Zu}}, \bibinfo {author} {\bibfnamefont {W.-B.}\ \bibnamefont {Wang}},
  \bibinfo {author} {\bibfnamefont {L.}~\bibnamefont {He}}, \bibinfo {author}
  {\bibfnamefont {W.-G.}\ \bibnamefont {Zhang}}, \bibinfo {author}
  {\bibfnamefont {C.-Y.}\ \bibnamefont {Dai}}, \bibinfo {author} {\bibfnamefont
  {F.}~\bibnamefont {Wang}},\ and\ \bibinfo {author} {\bibfnamefont {L.-M.}\
  \bibnamefont {Duan}},\ }\bibfield  {title} {\bibinfo {title} {{Experimental
  realization of universal geometric quantum gates with solid-state spins}},\
  }\href {https://doi.org/10.1038/nature13729} {\bibfield  {journal} {\bibinfo
  {journal} {Nature}\ }\textbf {\bibinfo {volume} {514}},\ \bibinfo {pages}
  {72} (\bibinfo {year} {2014})}\BibitemShut {NoStop}%
\bibitem [{\citenamefont {Arroyo-Camejo}\ \emph {et~al.}(2014)\citenamefont
  {Arroyo-Camejo}, \citenamefont {Lazariev}, \citenamefont {Hell},\ and\
  \citenamefont {Balasubramanian}}]{Arroyo-Camejo.2014}%
  \BibitemOpen
  \bibfield  {author} {\bibinfo {author} {\bibfnamefont {S.}~\bibnamefont
  {Arroyo-Camejo}}, \bibinfo {author} {\bibfnamefont {A.}~\bibnamefont
  {Lazariev}}, \bibinfo {author} {\bibfnamefont {S.~W.}\ \bibnamefont {Hell}},\
  and\ \bibinfo {author} {\bibfnamefont {G.}~\bibnamefont {Balasubramanian}},\
  }\bibfield  {title} {\bibinfo {title} {{Room temperature high-fidelity
  holonomic single-qubit gate on a solid-state spin}},\ }\href
  {https://doi.org/10.1038/ncomms5870} {\bibfield  {journal} {\bibinfo
  {journal} {Nature Communications}\ }\textbf {\bibinfo {volume} {5}},\
  \bibinfo {pages} {4870} (\bibinfo {year} {2014})}\BibitemShut {NoStop}%
\bibitem [{\citenamefont {Novi\v{c}enko}\ and\ \citenamefont
  {Juzeli\={u}nas}(2019)}]{Novičenko.2019}%
  \BibitemOpen
  \bibfield  {author} {\bibinfo {author} {\bibfnamefont {V.}~\bibnamefont
  {Novi\v{c}enko}}\ and\ \bibinfo {author} {\bibfnamefont {G.}~\bibnamefont
  {Juzeli\={u}nas}},\ }\bibfield  {title} {\bibinfo {title} {{Non-Abelian
  geometric phases in periodically driven systems}},\ }\href
  {https://doi.org/10.1103/physreva.100.012127} {\bibfield  {journal} {\bibinfo
   {journal} {Physical Review A}\ }\textbf {\bibinfo {volume} {100}},\ \bibinfo
  {pages} {012127} (\bibinfo {year} {2019})}\BibitemShut {NoStop}%
\bibitem [{\citenamefont {Novi\v{c}enko}\ \emph {et~al.}(2017)\citenamefont
  {Novi\v{c}enko}, \citenamefont {Anisimovas},\ and\ \citenamefont
  {Juzeli\={u}nas}}]{Novičenko.2017}%
  \BibitemOpen
  \bibfield  {author} {\bibinfo {author} {\bibfnamefont {V.}~\bibnamefont
  {Novi\v{c}enko}}, \bibinfo {author} {\bibfnamefont {E.}~\bibnamefont
  {Anisimovas}},\ and\ \bibinfo {author} {\bibfnamefont {G.}~\bibnamefont
  {Juzeli\={u}nas}},\ }\bibfield  {title} {\bibinfo {title} {{Floquet analysis
  of a quantum system with modulated periodic driving}},\ }\href
  {https://doi.org/10.1103/physreva.95.023615} {\bibfield  {journal} {\bibinfo
  {journal} {Physical Review A}\ }\textbf {\bibinfo {volume} {95}},\ \bibinfo
  {pages} {023615} (\bibinfo {year} {2017})}\BibitemShut {NoStop}%
\bibitem [{\citenamefont {Chen}\ \emph {et~al.}(2020)\citenamefont {Chen},
  \citenamefont {Murphree},\ and\ \citenamefont {Bigelow}}]{Chen.2020}%
  \BibitemOpen
  \bibfield  {author} {\bibinfo {author} {\bibfnamefont {Z.}~\bibnamefont
  {Chen}}, \bibinfo {author} {\bibfnamefont {J.~D.}\ \bibnamefont {Murphree}},\
  and\ \bibinfo {author} {\bibfnamefont {N.~P.}\ \bibnamefont {Bigelow}},\
  }\bibfield  {title} {\bibinfo {title} {{SU(2) geometric phase induced by a
  periodically driven Raman process in an ultracold dilute Bose gas}},\ }\href
  {https://doi.org/10.1103/physreva.101.013606} {\bibfield  {journal} {\bibinfo
   {journal} {Physical Review A}\ }\textbf {\bibinfo {volume} {101}},\ \bibinfo
  {pages} {013606} (\bibinfo {year} {2020})}\BibitemShut {NoStop}%
\bibitem [{\citenamefont {Wang}\ \emph {et~al.}(2021)\citenamefont {Wang},
  \citenamefont {Liu}, \citenamefont {Su},\ and\ \citenamefont
  {Yung}}]{Wang.2021}%
  \BibitemOpen
  \bibfield  {author} {\bibinfo {author} {\bibfnamefont {Y.-S.}\ \bibnamefont
  {Wang}}, \bibinfo {author} {\bibfnamefont {B.-J.}\ \bibnamefont {Liu}},
  \bibinfo {author} {\bibfnamefont {S.-L.}\ \bibnamefont {Su}},\ and\ \bibinfo
  {author} {\bibfnamefont {M.-H.}\ \bibnamefont {Yung}},\ }\bibfield  {title}
  {\bibinfo {title} {{Error-resilient Floquet geometric quantum computation}},\
  }\href {https://doi.org/10.1103/physrevresearch.3.033010} {\bibfield
  {journal} {\bibinfo  {journal} {Physical Review Research}\ }\textbf {\bibinfo
  {volume} {3}},\ \bibinfo {pages} {033010} (\bibinfo {year}
  {2021})}\BibitemShut {NoStop}%
\bibitem [{\citenamefont {Galitski}\ \emph {et~al.}(2019)\citenamefont
  {Galitski}, \citenamefont {Juzeli\={u}nas},\ and\ \citenamefont
  {Spielman}}]{Galitski.2019}%
  \BibitemOpen
  \bibfield  {author} {\bibinfo {author} {\bibfnamefont {V.}~\bibnamefont
  {Galitski}}, \bibinfo {author} {\bibfnamefont {G.}~\bibnamefont
  {Juzeli\={u}nas}},\ and\ \bibinfo {author} {\bibfnamefont {I.~B.}\
  \bibnamefont {Spielman}},\ }\bibfield  {title} {\bibinfo {title} {{Artificial
  gauge fields with ultracold atoms}},\ }\href
  {https://doi.org/10.1063/pt.3.4111} {\bibfield  {journal} {\bibinfo
  {journal} {Physics Today}\ }\textbf {\bibinfo {volume} {72}},\ \bibinfo
  {pages} {38} (\bibinfo {year} {2019})}\BibitemShut {NoStop}%
\bibitem [{\citenamefont {Sugawa}\ \emph {et~al.}(2021)\citenamefont {Sugawa},
  \citenamefont {Salces-Carcoba}, \citenamefont {Yue}, \citenamefont {Putra},\
  and\ \citenamefont {Spielman}}]{Sugawa.2021}%
  \BibitemOpen
  \bibfield  {author} {\bibinfo {author} {\bibfnamefont {S.}~\bibnamefont
  {Sugawa}}, \bibinfo {author} {\bibfnamefont {F.}~\bibnamefont
  {Salces-Carcoba}}, \bibinfo {author} {\bibfnamefont {Y.}~\bibnamefont {Yue}},
  \bibinfo {author} {\bibfnamefont {A.}~\bibnamefont {Putra}},\ and\ \bibinfo
  {author} {\bibfnamefont {I.~B.}\ \bibnamefont {Spielman}},\ }\bibfield
  {title} {\bibinfo {title} {{Wilson loop and Wilczek-Zee phase from a
  non-Abelian gauge field}},\ }\href
  {https://doi.org/10.1038/s41534-021-00483-2} {\bibfield  {journal} {\bibinfo
  {journal} {npj Quantum Information}\ }\textbf {\bibinfo {volume} {7}},\
  \bibinfo {pages} {144} (\bibinfo {year} {2021})}\BibitemShut {NoStop}%
\bibitem [{\citenamefont {Wilczek}\ and\ \citenamefont
  {Zee}(1984)}]{Wilczek.1984}%
  \BibitemOpen
  \bibfield  {author} {\bibinfo {author} {\bibfnamefont {F.}~\bibnamefont
  {Wilczek}}\ and\ \bibinfo {author} {\bibfnamefont {A.}~\bibnamefont {Zee}},\
  }\bibfield  {title} {\bibinfo {title} {{Appearance of Gauge Structure in
  Simple Dynamical Systems}},\ }\href
  {https://doi.org/10.1103/physrevlett.52.2111} {\bibfield  {journal} {\bibinfo
   {journal} {Physical Review Letters}\ }\textbf {\bibinfo {volume} {52}},\
  \bibinfo {pages} {2111} (\bibinfo {year} {1984})}\BibitemShut {NoStop}%
\bibitem [{\citenamefont {Anandan}(1988)}]{Anandan.1988}%
  \BibitemOpen
  \bibfield  {author} {\bibinfo {author} {\bibfnamefont {J.}~\bibnamefont
  {Anandan}},\ }\bibfield  {title} {\bibinfo {title} {{Non-adiabatic
  non-abelian geometric phase}},\ }\href
  {https://doi.org/10.1016/0375-9601(88)91010-9} {\bibfield  {journal}
  {\bibinfo  {journal} {Physics Letters A}\ }\textbf {\bibinfo {volume}
  {133}},\ \bibinfo {pages} {171} (\bibinfo {year} {1988})}\BibitemShut
  {NoStop}%
\bibitem [{\citenamefont {Karp}\ \emph {et~al.}(1999)\citenamefont {Karp},
  \citenamefont {Mansouri},\ and\ \citenamefont {Rno}}]{Karp.1999}%
  \BibitemOpen
  \bibfield  {author} {\bibinfo {author} {\bibfnamefont {R.~L.}\ \bibnamefont
  {Karp}}, \bibinfo {author} {\bibfnamefont {F.}~\bibnamefont {Mansouri}},\
  and\ \bibinfo {author} {\bibfnamefont {J.~S.}\ \bibnamefont {Rno}},\
  }\bibfield  {title} {\bibinfo {title} {{Product integral formalism and
  non-Abelian Stokes theorem}},\ }\href {https://doi.org/10.1063/1.533068}
  {\bibfield  {journal} {\bibinfo  {journal} {Journal of Mathematical Physics}\
  }\textbf {\bibinfo {volume} {40}},\ \bibinfo {pages} {6033} (\bibinfo {year}
  {1999})}\BibitemShut {NoStop}%
\bibitem [{\citenamefont {Viola}\ \emph {et~al.}(1999)\citenamefont {Viola},
  \citenamefont {Knill},\ and\ \citenamefont {Lloyd}}]{Viola.1999}%
  \BibitemOpen
  \bibfield  {author} {\bibinfo {author} {\bibfnamefont {L.}~\bibnamefont
  {Viola}}, \bibinfo {author} {\bibfnamefont {E.}~\bibnamefont {Knill}},\ and\
  \bibinfo {author} {\bibfnamefont {S.}~\bibnamefont {Lloyd}},\ }\bibfield
  {title} {\bibinfo {title} {{Dynamical Decoupling of Open Quantum Systems}},\
  }\href {https://doi.org/10.1103/physrevlett.82.2417} {\bibfield  {journal}
  {\bibinfo  {journal} {Physical Review Letters}\ }\textbf {\bibinfo {volume}
  {82}},\ \bibinfo {pages} {2417} (\bibinfo {year} {1999})}\BibitemShut
  {NoStop}%
\bibitem [{\citenamefont {Khodjasteh}\ and\ \citenamefont
  {Lidar}(2005)}]{Khodjasteh.2005}%
  \BibitemOpen
  \bibfield  {author} {\bibinfo {author} {\bibfnamefont {K.}~\bibnamefont
  {Khodjasteh}}\ and\ \bibinfo {author} {\bibfnamefont {D.~A.}\ \bibnamefont
  {Lidar}},\ }\bibfield  {title} {\bibinfo {title} {{Fault-Tolerant Quantum
  Dynamical Decoupling}},\ }\href
  {https://doi.org/10.1103/physrevlett.95.180501} {\bibfield  {journal}
  {\bibinfo  {journal} {Physical Review Letters}\ }\textbf {\bibinfo {volume}
  {95}},\ \bibinfo {pages} {180501} (\bibinfo {year} {2005})}\BibitemShut
  {NoStop}%
\bibitem [{\citenamefont {Zanardi}(1999)}]{Zanardi.1999}%
  \BibitemOpen
  \bibfield  {author} {\bibinfo {author} {\bibfnamefont {P.}~\bibnamefont
  {Zanardi}},\ }\bibfield  {title} {\bibinfo {title} {{Symmetrizing
  evolutions}},\ }\href {https://doi.org/10.1016/s0375-9601(99)00365-5}
  {\bibfield  {journal} {\bibinfo  {journal} {Physics Letters A}\ }\textbf
  {\bibinfo {volume} {258}},\ \bibinfo {pages} {77} (\bibinfo {year}
  {1999})}\BibitemShut {NoStop}%
\bibitem [{\citenamefont {Wu}\ and\ \citenamefont {Zhao}(2022)}]{Wu.2022h04}%
  \BibitemOpen
  \bibfield  {author} {\bibinfo {author} {\bibfnamefont {X.}~\bibnamefont
  {Wu}}\ and\ \bibinfo {author} {\bibfnamefont {P.~Z.}\ \bibnamefont {Zhao}},\
  }\bibfield  {title} {\bibinfo {title} {{Nonadiabatic geometric quantum
  computation protected by dynamical decoupling via the XXZ Hamiltonian}},\
  }\href {https://doi.org/10.1007/s11467-021-1128-z} {\bibfield  {journal}
  {\bibinfo  {journal} {Frontiers of Physics}\ }\textbf {\bibinfo {volume}
  {17}},\ \bibinfo {pages} {31502} (\bibinfo {year} {2022})}\BibitemShut
  {NoStop}%
\bibitem [{\citenamefont {Zhao}\ \emph {et~al.}(2021)\citenamefont {Zhao},
  \citenamefont {Wu},\ and\ \citenamefont {Tong}}]{Zhao.2021}%
  \BibitemOpen
  \bibfield  {author} {\bibinfo {author} {\bibfnamefont {P.~Z.}\ \bibnamefont
  {Zhao}}, \bibinfo {author} {\bibfnamefont {X.}~\bibnamefont {Wu}},\ and\
  \bibinfo {author} {\bibfnamefont {D.~M.}\ \bibnamefont {Tong}},\ }\bibfield
  {title} {\bibinfo {title} {{Dynamical-decoupling-protected nonadiabatic
  holonomic quantum computation}},\ }\href
  {https://doi.org/10.1103/physreva.103.012205} {\bibfield  {journal} {\bibinfo
   {journal} {Physical Review A}\ }\textbf {\bibinfo {volume} {103}},\ \bibinfo
  {pages} {012205} (\bibinfo {year} {2021})}\BibitemShut {NoStop}%
\bibitem [{\citenamefont {Liang}\ and\ \citenamefont {Xue}(2022)}]{Liang.2022}%
  \BibitemOpen
  \bibfield  {author} {\bibinfo {author} {\bibfnamefont {M.-J.}\ \bibnamefont
  {Liang}}\ and\ \bibinfo {author} {\bibfnamefont {Z.-Y.}\ \bibnamefont
  {Xue}},\ }\bibfield  {title} {\bibinfo {title} {{Robust nonadiabatic
  geometric quantum computation by dynamical correction}},\ }\href
  {https://doi.org/10.1103/physreva.106.012603} {\bibfield  {journal} {\bibinfo
   {journal} {Physical Review A}\ }\textbf {\bibinfo {volume} {106}},\ \bibinfo
  {pages} {012603} (\bibinfo {year} {2022})}\BibitemShut {NoStop}%
\bibitem [{\citenamefont {Lindon}\ \emph {et~al.}(2023)\citenamefont {Lindon},
  \citenamefont {Tashchilina}, \citenamefont {Cooke},\ and\ \citenamefont
  {LeBlanc}}]{Lindon.2023}%
  \BibitemOpen
  \bibfield  {author} {\bibinfo {author} {\bibfnamefont {J.}~\bibnamefont
  {Lindon}}, \bibinfo {author} {\bibfnamefont {A.}~\bibnamefont {Tashchilina}},
  \bibinfo {author} {\bibfnamefont {L.~W.}\ \bibnamefont {Cooke}},\ and\
  \bibinfo {author} {\bibfnamefont {L.~J.}\ \bibnamefont {LeBlanc}},\
  }\bibfield  {title} {\bibinfo {title} {{Complete Unitary Qutrit Control in
  Ultracold Atoms}},\ }\href {https://doi.org/10.1103/physrevapplied.19.034089}
  {\bibfield  {journal} {\bibinfo  {journal} {Physical Review Applied}\
  }\textbf {\bibinfo {volume} {19}},\ \bibinfo {pages} {034089} (\bibinfo
  {year} {2023})}\BibitemShut {NoStop}%
\bibitem [{\citenamefont {Das}(2018)}]{Das.2018}%
  \BibitemOpen
  \bibfield  {author} {\bibinfo {author} {\bibfnamefont {K.~K.}\ \bibnamefont
  {Das}},\ }\bibfield  {title} {\bibinfo {title} {{Measurement and significance
  of Wilson loops in synthetic gauge fields}},\ }\href
  {https://doi.org/10.1103/physreva.97.053620} {\bibfield  {journal} {\bibinfo
  {journal} {Physical Review A}\ }\textbf {\bibinfo {volume} {97}},\ \bibinfo
  {pages} {053620} (\bibinfo {year} {2018})}\BibitemShut {NoStop}%
\bibitem [{\citenamefont {Aidelsburger}\ \emph {et~al.}(2018)\citenamefont
  {Aidelsburger}, \citenamefont {Nascimbene},\ and\ \citenamefont
  {Goldman}}]{Aidelsburger.2018}%
  \BibitemOpen
  \bibfield  {author} {\bibinfo {author} {\bibfnamefont {M.}~\bibnamefont
  {Aidelsburger}}, \bibinfo {author} {\bibfnamefont {S.}~\bibnamefont
  {Nascimbene}},\ and\ \bibinfo {author} {\bibfnamefont {N.}~\bibnamefont
  {Goldman}},\ }\bibfield  {title} {\bibinfo {title} {{Artificial gauge fields
  in materials and engineered systems}},\ }\href
  {https://doi.org/10.1016/j.crhy.2018.03.002} {\bibfield  {journal} {\bibinfo
  {journal} {Comptes Rendus Physique}\ }\textbf {\bibinfo {volume} {19}},\
  \bibinfo {pages} {394} (\bibinfo {year} {2018})}\BibitemShut {NoStop}%
\bibitem [{\citenamefont {Zee}(1988)}]{Zee.1988}%
  \BibitemOpen
  \bibfield  {author} {\bibinfo {author} {\bibfnamefont {A.}~\bibnamefont
  {Zee}},\ }\bibfield  {title} {\bibinfo {title} {{Non-Abelian gauge structure
  in nuclear quadrupole resonance}},\ }\href
  {https://doi.org/10.1103/physreva.38.1} {\bibfield  {journal} {\bibinfo
  {journal} {Physical Review A}\ }\textbf {\bibinfo {volume} {38}},\ \bibinfo
  {pages} {1} (\bibinfo {year} {1988})}\BibitemShut {NoStop}%
\bibitem [{\citenamefont {Budker}\ and\ \citenamefont
  {Romalis}(2007)}]{Budker.2007}%
  \BibitemOpen
  \bibfield  {author} {\bibinfo {author} {\bibfnamefont {D.}~\bibnamefont
  {Budker}}\ and\ \bibinfo {author} {\bibfnamefont {M.}~\bibnamefont
  {Romalis}},\ }\bibfield  {title} {\bibinfo {title} {{Optical magnetometry}},\
  }\href {https://doi.org/10.1038/nphys566} {\bibfield  {journal} {\bibinfo
  {journal} {Nature Physics}\ }\textbf {\bibinfo {volume} {3}},\ \bibinfo
  {pages} {227} (\bibinfo {year} {2007})}\BibitemShut {NoStop}%
\bibitem [{\citenamefont {Borkowski}\ \emph {et~al.}(2023)\citenamefont
  {Borkowski}, \citenamefont {Reichs\"{o}llner}, \citenamefont {Thekkeppatt},
  \citenamefont {Barb\'{e}}, \citenamefont {Roon}, \citenamefont {Druten},\
  and\ \citenamefont {Schreck}}]{Borkowski.2023}%
  \BibitemOpen
  \bibfield  {author} {\bibinfo {author} {\bibfnamefont {M.}~\bibnamefont
  {Borkowski}}, \bibinfo {author} {\bibfnamefont {L.}~\bibnamefont
  {Reichs\"{o}llner}}, \bibinfo {author} {\bibfnamefont {P.}~\bibnamefont
  {Thekkeppatt}}, \bibinfo {author} {\bibfnamefont {V.}~\bibnamefont
  {Barb\'{e}}}, \bibinfo {author} {\bibfnamefont {T.~v.}\ \bibnamefont {Roon}},
  \bibinfo {author} {\bibfnamefont {K.~v.}\ \bibnamefont {Druten}},\ and\
  \bibinfo {author} {\bibfnamefont {F.}~\bibnamefont {Schreck}},\ }\bibfield
  {title} {\bibinfo {title} {{Active stabilization of kilogauss magnetic fields
  to the ppm level for magnetoassociation on ultranarrow Feshbach
  resonances}},\ }\href {https://doi.org/10.1063/5.0143825} {\bibfield
  {journal} {\bibinfo  {journal} {Review of Scientific Instruments}\ }\textbf
  {\bibinfo {volume} {94}},\ \bibinfo {pages} {073202} (\bibinfo {year}
  {2023})}\BibitemShut {NoStop}%
\bibitem [{\citenamefont {Xu}\ \emph {et~al.}(2019)\citenamefont {Xu},
  \citenamefont {Wang}, \citenamefont {Jiao}, \citenamefont {Yi}, \citenamefont
  {Sun},\ and\ \citenamefont {Chen}}]{Xu.2019}%
  \BibitemOpen
  \bibfield  {author} {\bibinfo {author} {\bibfnamefont {X.-T.}\ \bibnamefont
  {Xu}}, \bibinfo {author} {\bibfnamefont {Z.-Y.}\ \bibnamefont {Wang}},
  \bibinfo {author} {\bibfnamefont {R.-H.}\ \bibnamefont {Jiao}}, \bibinfo
  {author} {\bibfnamefont {C.-R.}\ \bibnamefont {Yi}}, \bibinfo {author}
  {\bibfnamefont {W.}~\bibnamefont {Sun}},\ and\ \bibinfo {author}
  {\bibfnamefont {S.}~\bibnamefont {Chen}},\ }\bibfield  {title} {\bibinfo
  {title} {{Ultra-low noise magnetic field for quantum gases}},\ }\href
  {https://doi.org/10.1063/1.5087957} {\bibfield  {journal} {\bibinfo
  {journal} {Review of Scientific Instruments}\ }\textbf {\bibinfo {volume}
  {90}},\ \bibinfo {pages} {054708} (\bibinfo {year} {2019})}\BibitemShut
  {NoStop}%
\bibitem [{\citenamefont {Lahtinen}\ and\ \citenamefont
  {Pachos}(2017)}]{Lahtinen.2017}%
  \BibitemOpen
  \bibfield  {author} {\bibinfo {author} {\bibfnamefont {V.}~\bibnamefont
  {Lahtinen}}\ and\ \bibinfo {author} {\bibfnamefont {J.}~\bibnamefont
  {Pachos}},\ }\bibfield  {title} {\bibinfo {title} {{A Short Introduction to
  Topological Quantum Computation}},\ }\href
  {https://doi.org/10.21468/scipostphys.3.3.021} {\bibfield  {journal}
  {\bibinfo  {journal} {SciPost Physics}\ }\textbf {\bibinfo {volume} {3}},\
  \bibinfo {pages} {021} (\bibinfo {year} {2017})}\BibitemShut {NoStop}%
\bibitem [{\citenamefont {Nayak}\ \emph {et~al.}(2008)\citenamefont {Nayak},
  \citenamefont {Simon}, \citenamefont {Stern}, \citenamefont {Freedman},\ and\
  \citenamefont {Das~Sarma}}]{Nayak.2008}%
  \BibitemOpen
  \bibfield  {author} {\bibinfo {author} {\bibfnamefont {C.}~\bibnamefont
  {Nayak}}, \bibinfo {author} {\bibfnamefont {S.~H.}\ \bibnamefont {Simon}},
  \bibinfo {author} {\bibfnamefont {A.}~\bibnamefont {Stern}}, \bibinfo
  {author} {\bibfnamefont {M.}~\bibnamefont {Freedman}},\ and\ \bibinfo
  {author} {\bibfnamefont {S.}~\bibnamefont {Das~Sarma}},\ }\bibfield  {title}
  {\bibinfo {title} {{Non-Abelian anyons and topological quantum
  computation}},\ }\href {https://doi.org/10.1103/revmodphys.80.1083}
  {\bibfield  {journal} {\bibinfo  {journal} {Reviews of Modern Physics}\
  }\textbf {\bibinfo {volume} {80}},\ \bibinfo {pages} {1083} (\bibinfo {year}
  {2008})}\BibitemShut {NoStop}%
\bibitem [{\citenamefont {Li}\ \emph {et~al.}(2021)\citenamefont {Li},
  \citenamefont {Xu},\ and\ \citenamefont {Tong}}]{Li.2021e63}%
  \BibitemOpen
  \bibfield  {author} {\bibinfo {author} {\bibfnamefont {K.~Z.}\ \bibnamefont
  {Li}}, \bibinfo {author} {\bibfnamefont {G.~F.}\ \bibnamefont {Xu}},\ and\
  \bibinfo {author} {\bibfnamefont {D.~M.}\ \bibnamefont {Tong}},\ }\bibfield
  {title} {\bibinfo {title} {{Coherence-protected nonadiabatic geometric
  quantum computation}},\ }\href
  {https://doi.org/10.1103/physrevresearch.3.023104} {\bibfield  {journal}
  {\bibinfo  {journal} {Physical Review Research}\ }\textbf {\bibinfo {volume}
  {3}},\ \bibinfo {pages} {023104} (\bibinfo {year} {2021})}\BibitemShut
  {NoStop}%
\bibitem [{\citenamefont {Liu}\ \emph {et~al.}(2023)\citenamefont {Liu},
  \citenamefont {Yan}, \citenamefont {Zhang}, \citenamefont {Yung},
  \citenamefont {Su},\ and\ \citenamefont {Shan}}]{Liu.2023gd}%
  \BibitemOpen
  \bibfield  {author} {\bibinfo {author} {\bibfnamefont {B.-J.}\ \bibnamefont
  {Liu}}, \bibinfo {author} {\bibfnamefont {L.-L.}\ \bibnamefont {Yan}},
  \bibinfo {author} {\bibfnamefont {Y.}~\bibnamefont {Zhang}}, \bibinfo
  {author} {\bibfnamefont {M.-H.}\ \bibnamefont {Yung}}, \bibinfo {author}
  {\bibfnamefont {S.-L.}\ \bibnamefont {Su}},\ and\ \bibinfo {author}
  {\bibfnamefont {C.-X.}\ \bibnamefont {Shan}},\ }\bibfield  {title} {\bibinfo
  {title} {{Decoherence-suppressed nonadiabatic holonomic quantum
  computation}},\ }\href {https://doi.org/10.1103/physrevresearch.5.013059}
  {\bibfield  {journal} {\bibinfo  {journal} {Physical Review Research}\
  }\textbf {\bibinfo {volume} {5}},\ \bibinfo {pages} {013059} (\bibinfo {year}
  {2023})}\BibitemShut {NoStop}%
\bibitem [{\citenamefont {Kang}\ \emph {et~al.}(2022)\citenamefont {Kang},
  \citenamefont {Chen}, \citenamefont {Wang}, \citenamefont {Song},
  \citenamefont {Xia}, \citenamefont {Miranowicz}, \citenamefont {Zheng},\ and\
  \citenamefont {Nori}}]{Kang.202272u}%
  \BibitemOpen
  \bibfield  {author} {\bibinfo {author} {\bibfnamefont {Y.-H.}\ \bibnamefont
  {Kang}}, \bibinfo {author} {\bibfnamefont {Y.-H.}\ \bibnamefont {Chen}},
  \bibinfo {author} {\bibfnamefont {X.}~\bibnamefont {Wang}}, \bibinfo {author}
  {\bibfnamefont {J.}~\bibnamefont {Song}}, \bibinfo {author} {\bibfnamefont
  {Y.}~\bibnamefont {Xia}}, \bibinfo {author} {\bibfnamefont {A.}~\bibnamefont
  {Miranowicz}}, \bibinfo {author} {\bibfnamefont {S.-B.}\ \bibnamefont
  {Zheng}},\ and\ \bibinfo {author} {\bibfnamefont {F.}~\bibnamefont {Nori}},\
  }\bibfield  {title} {\bibinfo {title} {{Nonadiabatic geometric quantum
  computation with cat-state qubits via invariant-based reverse engineering}},\
  }\href {https://doi.org/10.1103/physrevresearch.4.013233} {\bibfield
  {journal} {\bibinfo  {journal} {Physical Review Research}\ }\textbf {\bibinfo
  {volume} {4}},\ \bibinfo {pages} {013233} (\bibinfo {year}
  {2022})}\BibitemShut {NoStop}%
\bibitem [{\citenamefont {Xu}\ \emph {et~al.}(2012)\citenamefont {Xu},
  \citenamefont {Zhang}, \citenamefont {Tong}, \citenamefont {Sj\"{o}qvist},\
  and\ \citenamefont {Kwek}}]{Xu.2012}%
  \BibitemOpen
  \bibfield  {author} {\bibinfo {author} {\bibfnamefont {G.~F.}\ \bibnamefont
  {Xu}}, \bibinfo {author} {\bibfnamefont {J.}~\bibnamefont {Zhang}}, \bibinfo
  {author} {\bibfnamefont {D.~M.}\ \bibnamefont {Tong}}, \bibinfo {author}
  {\bibfnamefont {E.}~\bibnamefont {Sj\"{o}qvist}},\ and\ \bibinfo {author}
  {\bibfnamefont {L.~C.}\ \bibnamefont {Kwek}},\ }\bibfield  {title} {\bibinfo
  {title} {{Nonadiabatic holonomic quantum computation in decoherence-free
  subspaces}},\ }\href {https://doi.org/10.1103/physrevlett.109.170501}
  {\bibfield  {journal} {\bibinfo  {journal} {Physical Review Letters}\
  }\textbf {\bibinfo {volume} {109}},\ \bibinfo {pages} {170501} (\bibinfo
  {year} {2012})}\BibitemShut {NoStop}%
\bibitem [{\citenamefont {Oreshkov}\ \emph {et~al.}(2009)\citenamefont
  {Oreshkov}, \citenamefont {Brun},\ and\ \citenamefont
  {Lidar}}]{Oreshkov.2009}%
  \BibitemOpen
  \bibfield  {author} {\bibinfo {author} {\bibfnamefont {O.}~\bibnamefont
  {Oreshkov}}, \bibinfo {author} {\bibfnamefont {T.~A.}\ \bibnamefont {Brun}},\
  and\ \bibinfo {author} {\bibfnamefont {D.~A.}\ \bibnamefont {Lidar}},\
  }\bibfield  {title} {\bibinfo {title} {Fault-tolerant holonomic quantum
  computation},\ }\href {https://doi.org/10.1103/PhysRevLett.102.070502}
  {\bibfield  {journal} {\bibinfo  {journal} {Phys. Rev. Lett.}\ }\textbf
  {\bibinfo {volume} {102}},\ \bibinfo {pages} {070502} (\bibinfo {year}
  {2009})}\BibitemShut {NoStop}%
\bibitem [{\citenamefont {Kang}\ \emph {et~al.}(2023)\citenamefont {Kang},
  \citenamefont {Xiao}, \citenamefont {Shi}, \citenamefont {Wang},
  \citenamefont {Yang}, \citenamefont {Song},\ and\ \citenamefont
  {Xia}}]{Kang.2023}%
  \BibitemOpen
  \bibfield  {author} {\bibinfo {author} {\bibfnamefont {Y.-H.}\ \bibnamefont
  {Kang}}, \bibinfo {author} {\bibfnamefont {Y.}~\bibnamefont {Xiao}}, \bibinfo
  {author} {\bibfnamefont {Z.-C.}\ \bibnamefont {Shi}}, \bibinfo {author}
  {\bibfnamefont {Y.}~\bibnamefont {Wang}}, \bibinfo {author} {\bibfnamefont
  {J.-Q.}\ \bibnamefont {Yang}}, \bibinfo {author} {\bibfnamefont
  {J.}~\bibnamefont {Song}},\ and\ \bibinfo {author} {\bibfnamefont
  {Y.}~\bibnamefont {Xia}},\ }\bibfield  {title} {\bibinfo {title} {{Effective
  implementation of nonadiabatic geometric quantum gates of cat-state qubits
  using an auxiliary qutrit}},\ }\href
  {https://doi.org/10.1088/1367-2630/acc2de} {\bibfield  {journal} {\bibinfo
  {journal} {New Journal of Physics}\ }\textbf {\bibinfo {volume} {25}},\
  \bibinfo {pages} {033029} (\bibinfo {year} {2023})}\BibitemShut {NoStop}%
\bibitem [{\citenamefont {Sakurai}\ and\ \citenamefont
  {Napolitano}(2021)}]{sakurai2014modern}%
  \BibitemOpen
  \bibfield  {author} {\bibinfo {author} {\bibfnamefont {J.~J.~S.}\
  \bibnamefont {Sakurai}}\ and\ \bibinfo {author} {\bibfnamefont {J.~J.}\
  \bibnamefont {Napolitano}},\ }\href@noop {} {\emph {\bibinfo {title} {Modern
  Quantum Mechanics}}}\ (\bibinfo  {publisher} {Cambridge},\ \bibinfo {year}
  {2021})\ Chap.\ \bibinfo {chapter} {3.9.2}\BibitemShut {NoStop}%
\bibitem [{\citenamefont {Jim\'{e}nez-Garc\'{i}a}\ \emph
  {et~al.}(2012)\citenamefont {Jim\'{e}nez-Garc\'{i}a}, \citenamefont
  {LeBlanc}, \citenamefont {Williams}, \citenamefont {Beeler}, \citenamefont
  {Perry},\ and\ \citenamefont {Spielman}}]{Jiménez-García.2012}%
  \BibitemOpen
  \bibfield  {author} {\bibinfo {author} {\bibfnamefont {K.}~\bibnamefont
  {Jim\'{e}nez-Garc\'{i}a}}, \bibinfo {author} {\bibfnamefont {L.~J.}\
  \bibnamefont {LeBlanc}}, \bibinfo {author} {\bibfnamefont {R.~A.}\
  \bibnamefont {Williams}}, \bibinfo {author} {\bibfnamefont {M.~C.}\
  \bibnamefont {Beeler}}, \bibinfo {author} {\bibfnamefont {A.~R.}\
  \bibnamefont {Perry}},\ and\ \bibinfo {author} {\bibfnamefont {I.~B.}\
  \bibnamefont {Spielman}},\ }\bibfield  {title} {\bibinfo {title} {{Peierls
  Substitution in an Engineered Lattice Potential}},\ }\href
  {https://doi.org/10.1103/physrevlett.108.225303} {\bibfield  {journal}
  {\bibinfo  {journal} {Physical Review Letters}\ }\textbf {\bibinfo {volume}
  {108}},\ \bibinfo {pages} {225303} (\bibinfo {year} {2012})}\BibitemShut
  {NoStop}%
\bibitem [{\citenamefont {Stamper-Kurn}\ and\ \citenamefont
  {Ueda}(2013)}]{Stamper-Kurn.2013}%
  \BibitemOpen
  \bibfield  {author} {\bibinfo {author} {\bibfnamefont {D.~M.}\ \bibnamefont
  {Stamper-Kurn}}\ and\ \bibinfo {author} {\bibfnamefont {M.}~\bibnamefont
  {Ueda}},\ }\bibfield  {title} {\bibinfo {title} {{Spinor Bose gases:
  Symmetries, magnetism, and quantum dynamics}},\ }\href
  {https://doi.org/10.1103/revmodphys.85.1191} {\bibfield  {journal} {\bibinfo
  {journal} {Reviews of Modern Physics}\ }\textbf {\bibinfo {volume} {85}},\
  \bibinfo {pages} {1191} (\bibinfo {year} {2013})}\BibitemShut {NoStop}%
\bibitem [{\citenamefont {Bezanson}\ \emph {et~al.}(2017)\citenamefont
  {Bezanson}, \citenamefont {Edelman}, \citenamefont {Karpinski},\ and\
  \citenamefont {Shah}}]{Julia-2017}%
  \BibitemOpen
  \bibfield  {author} {\bibinfo {author} {\bibfnamefont {J.}~\bibnamefont
  {Bezanson}}, \bibinfo {author} {\bibfnamefont {A.}~\bibnamefont {Edelman}},
  \bibinfo {author} {\bibfnamefont {S.}~\bibnamefont {Karpinski}},\ and\
  \bibinfo {author} {\bibfnamefont {V.~B.}\ \bibnamefont {Shah}},\ }\bibfield
  {title} {\bibinfo {title} {Julia: A fresh approach to numerical computing},\
  }\href {https://doi.org/10.1137/141000671} {\bibfield  {journal} {\bibinfo
  {journal} {SIAM {R}eview}\ }\textbf {\bibinfo {volume} {59}},\ \bibinfo
  {pages} {65} (\bibinfo {year} {2017})}\BibitemShut {NoStop}%
\bibitem [{\citenamefont {Rackauckas}\ and\ \citenamefont
  {Nie}(2017)}]{DifferentialEquations.jl-2017}%
  \BibitemOpen
  \bibfield  {author} {\bibinfo {author} {\bibfnamefont {C.}~\bibnamefont
  {Rackauckas}}\ and\ \bibinfo {author} {\bibfnamefont {Q.}~\bibnamefont
  {Nie}},\ }\bibfield  {title} {\bibinfo {title} {Differentialequations.jl –
  a performant and feature-rich ecosystem for solving differential equations in
  julia},\ }\href {https://doi.org/10.5334/jors.151} {\bibfield  {journal}
  {\bibinfo  {journal} {The Journal of Open Research Software}\ }\textbf
  {\bibinfo {volume} {5}} (\bibinfo {year} {2017})},\ \bibinfo {note} {exported
  from https://app.dimensions.ai on 2019/05/05}\BibitemShut {NoStop}%
\bibitem [{\citenamefont {Mogensen}\ and\ \citenamefont
  {Riseth}(2018)}]{Optim.jl-2018}%
  \BibitemOpen
  \bibfield  {author} {\bibinfo {author} {\bibfnamefont {P.~K.}\ \bibnamefont
  {Mogensen}}\ and\ \bibinfo {author} {\bibfnamefont {A.~N.}\ \bibnamefont
  {Riseth}},\ }\bibfield  {title} {\bibinfo {title} {Optim: A mathematical
  optimization package for {Julia}},\ }\href
  {https://doi.org/10.21105/joss.00615} {\bibfield  {journal} {\bibinfo
  {journal} {Journal of Open Source Software}\ }\textbf {\bibinfo {volume}
  {3}},\ \bibinfo {pages} {615} (\bibinfo {year} {2018})}\BibitemShut {NoStop}%
\end{thebibliography}
\end{document}